\documentclass[aps,prd,preprint,tightening,showpacs]{revtex4-2}
\usepackage{amsmath,amssymb,amsthm,graphicx}
\usepackage{upgreek}
\usepackage[mathscr]{eucal}
\usepackage[colorlinks,linkcolor=blue,
anchorcolor=blue,citecolor=green]{hyperref}
\usepackage[dvipsnames,usenames]{color}
\newcommand{\no}{\nonumber}
\newcommand{\be}{\begin{equation}}
\newcommand{\ee}{\end{equation}}
\newcommand{\ba}{\begin{eqnarray}}

\newcommand{\ea}{\end{eqnarray}}
\usepackage{epsf,epsfig,graphics}
\usepackage{verbatim,color,ulem}
\DeclareRobustCommand{\varlambda}{\text{\usefont{OML}{txmi}{m}{it}\symbol{"15}}}

\begin{document}
\title{Null geodesics in extremal Kerr-Newman black holes}
\author{Bo-Ruei Chen}
\author{Tien Hsieh}
\author{Da-Shin Lee}
\email{dslee@gms.ndhu.edu.tw}
\affiliation{
Department of Physics, National Dong Hwa University, Hualien, Taiwan, Republic of China}
\date{\today}

\begin{abstract}
We study the null geodesics in the extremal Kerr-Newman exterior.
We clarify the roots of the radial potential and obtain the parameter space of the azimuthal angular momentum and the Carter constant of the light rays for varieties of the orbits.
It is known that one of the unique features of extremal black holes for the null geodesics is the existence of the stable double root at the horizon, giving rise to the stable spherical motion.
For the black hole's spin $a<M/2$, the stable double root is isolated from the unstable one.
However, for $ a\ge M/2$, the unstable and stable double roots merge at the triple root so that the unstable double root in some parameter region can lie at the horizon, giving a very different shape to the light ring.
We then find the analytical expressions of light orbits, which can reach spatial infinity for both nonequatorial and equatorial motions.
In particular, for the orbits starting from the near horizon of the extremal Kerr-Newman black holes with the parameters for the unstable double and triple roots, the solutions are remarkably simple in terms of elementary functions.
It is also found that the analytical solutions of the equatorial motion can shed light on the deflection of the light by black holes.
Varying the azimuthal angular momentum, as either the double or triple root at the horizon is approached from the turning point, the stronger power-law divergence in the deflection angle is found in comparison with the typical logarithmic divergence in nonextremal black holes in the strong deflection limit.
This could be another interesting effect of light deflection by extremal black holes.
\end{abstract}

\pacs{04.70.-s, 04.70.Bw, 04.80.Cc}

\maketitle
\newpage

\section{Introduction}
The Event Horizon Telescope team has succeeded in the observation of the black hole shadows of the supermassive black holes M87* at the center of the M87 galaxy \cite{akiyama-2019} and Sagittarius A* (Sgr A*) at the center of our Galaxy \cite{collaboration-2022}.
The light ring is a shell of light that surrounds the shadow of the black hole, allowing the light to travel from the edge of the ring toward us.
The light can either eventually fall into the black hole and never return, or it can escape the orbit and travel to an observer in spatial infinity.
The brightness around the shadow certainly depends on how the light rays escape to reach spatial infinity.

Maximally spinning or charged black holes are known as extremal black holes.
Unlike nonextremal black holes, the extremal black holes have their intrinsic theoretical problems, such as those related to black hole thermodynamics.
One of these is that it is not possible to make nonextremal black holes become extremal in a finite amount of time by any continuous process \cite{israel-1986}.
In \cite{aretakis-2015}, the extremal Kerr-Newman black holes suffer from horizon instabilities.
In addition, there are stable light orbits at the horizons for extremal Kerr-Newman black holes \cite{ulbricht-2015, dolan-2016, khoo-2016}, which can be associated with distinct phenomenological features, including the trapping and storage of electromagnetic energy in bound regions and allowing various instabilities \cite{dolan-2016}.
However, rapidly rotating black holes are of great interest because they can accelerate the particle to high energy as particle accelerators \cite{banados-2009, harada-2014}.
The emission of light from near the horizon in extremal Kerr black holes has been studied in \cite{porfyriadis-2017, gralla-2018}.
According to \cite{porfyriadis-2017}, the analytical expressions of the orbits may be useful for the study of a variety of the problems related to the observations of electromagnetic radiation, which originates from or passes through the near horizon \cite{laor-1991, beckwith-2004, dovciak-2004, brenneman-2006, dauser-2010}.
It is also hoped that the solutions can be applied to other potential observations such as extremal black hole shadows or orbiting hot spots \cite{gralla-2018}.
As mentioned in \cite{dolan-2016} and references therein, in the case of horizonless ultracompact boson stars \cite{cardoso-2014}, outer unstable light orbits are accompanied by inner stable light orbits, as in the case of extremal Kerr-Newman black holes.
The study of the null geodesics in the exterior of extremal Kerr-Newman black holes may shed light on the role of stable light orbits in these ultracompact stars.

In this paper, we focus on the null geodesics in extremal Kerr-Newman black holes.
Theoretical considerations, together with the recent observations of the structures near Sgr A* by the GRAVITY experiment \cite{abuter-2018}, indicate the possibility of the presence of a small electric charge of a central supermassive black hole \cite{zajacek-2018, zajacek-2019}.
Here we would like to extend the work of \cite{porfyriadis-2017, gralla-2018} to consider extremal Kerr-Newman black holes. The bending of the light and its time delay by Kerr-Newman black holes were explored in \cite{hsiao-2020, hsieh-2021A, hsieh-2021B, hsieh-2024}.
The null geodesics in the exterior of Kerr-Newman black holes were also examined in \cite{wang-2022}, where the equations of motion are known as a set of first order differential equations in Mino time.
In this paper, we will adopt the approach of \cite{wang-2022}, classifying the roots of radial and angular potentials and finding the analytical solutions of the orbits, among which we will focus on light rays that can reach spatial infinity and, in particular, the orbits leaving the near horizon of the extremal Kerr-Newman black holes (NHEKN).
These orbits are unique in extremal black holes and can carry information about the horizon to be observed \cite{ogasawara-2020}.
Another observational effect, which can be implied from the analytical solutions, is the deflection of the light by extremal black holes.
As will be seen, the corresponding deflection angle diverges as the turning point approaches the double or triple root of the radial potential in the strong deflection limit (SDL) \cite{bozza-2002, eiroa-2002, tsukamoto-2017}.
The stronger divergence will be seen in the extremal black hole in comparison to the corresponding roots in the nonextremal black hole.

The layout of the paper is as follows. In Sec. \ref{secII}, a minireview of the null geodesics is provided along with three conserved quantities of light orbits, which are the energy, azimuthal angular momentum, and the Carter constant.
The equations of motion can be recast into the integral forms involving two effective potentials.
In Sec. \ref{secIII}, we explore the roots of the radial potential, which in turn determine the types of the orbits.
In particular, the light rays with the parameters giving the unstable double root of the radial potential become important for the construction of the light ring of the black holes.
In Sec. \ref{secIV}, the analytical solutions of the orbits, which can reach spatial infinity, are all obtained.
Their implication in the deflection of the light will be explored.
In Sec. \ref{secV}, the conclusions are drawn.
For clarity of the notation and completeness of the paper, Appendixes \ref{appA} and \ref{appB} provide some of the relevant formulas derived in the earlier publications \cite{wang-2022}.
Appendix \ref{appC} summarizes the solution of the light orbits  with the parameters at point F1 or F2 shown in Fig. \ref{eta_lambda_2}, where the radial potential has a complex-conjugated pair and two real-valued roots.
Since in this case there is no turning point in the black hole exterior, the light rays will plunge directly into the black hole or escape to spatial infinity.

\section{Equations of motion for null geodesics}
\label{secII}
In this section, we provide a short review of the equations of motion for the light in the Kerr-Newman black hole exterior.
In the Boyer-Lindquist coordinates $(t,r,\theta,\phi)$,
the line element of the exterior of the Kerr-Newman black hole with the gravitational mass $M$, charge $Q$, and angular momentum per unit mass $a=J/M$ is described by
\begin{align}
ds^2 = -\frac{\Delta}{\Sigma}\left(dt-a \sin^2\theta d\phi \right)^2
+\frac{\sin^2\theta}{\Sigma}\left[(r^2+a^2)d\phi-adt \right]^2
+\frac{\Sigma}{\Delta}dr^2
+\Sigma d\theta^2 ,
\end{align}
where
\be
\begin{split}
\Sigma &=r^2+a^2\cos^2\theta, \\
\Delta &=r^2-2Mr+a^2+Q^2. \\
\end{split}
\ee
The roots of $\Delta (r)$ determine the outer/inner event horizons $r_{+}/r_{-}$ as
\be
r_{\pm} =M \pm \sqrt{M^2 -(a^2+Q^2)}\;
\ee
in general with the condition $0 \le a^2 +Q^2 \le M^2$.
For the extremal black holes, $a^2+Q^2=M^2$ is considered, where $\Delta=(r-r_h)^2$ with $r_{\pm}=r_h=M$.
For the asymptotically flat, stationary, and axial-symmetric black hole with the metric of the independence of $t$ and $\phi$,
there exist conserved quantities along the geodesics, which are the energy $E$ and the azimuthal angular momentum $L$ of the light orbits.
They can be constructed through the 4-velocity $u^\mu = {dx^\mu}/{d\sigma}$ in terms of the affine parameter $\sigma$ given by
\begin{align}
E \equiv -\xi^\mu_t u_\mu, \quad
L \equiv \xi^\mu_\phi u_\mu \,
\end{align}
with the associated Killing vectors
\begin{align}
\begin{aligned}
\xi^\mu_t &= \delta^\mu_t, \quad
\xi^\mu_\phi =\delta^\mu_\phi \, .
\end{aligned}
\end{align}
Additionally, another conserved quantity is the Carter constant explicitly obtained by
\begin{equation}
C= \Sigma^2\left(u^{\theta}\right)^2-a^2 {E}^2 \cos^2\theta +L^2\cot^2\theta\, .
\end{equation}
The null geodesics, $u^\mu u_\mu=0$, gives the equations of motion,
\begin{align}
&\frac{\Sigma}{E}\frac{d{r}}{d\sigma}=\pm_r\sqrt{R({r})} \, ,\label{r_eq_photon}\\
&\frac{\Sigma}{E}\frac{d\theta}{d\sigma}=\pm_{\theta}\sqrt{\Theta(\theta)} \, , \label{theta_eq_photon}\\
&\frac{\Sigma}{E}\frac{d\phi}{d\sigma}=\frac{{a}}{{\Delta}}\left({r}^2+{a}^2-{a}\lambda\right)-\frac{1}{\sin^{2}\theta}\left({a}\sin^2\theta-\lambda\right) \, , \label{phi_eq_photon}\\
&\frac{\Sigma}{E}\frac{d{t}}{d\sigma}=\frac{{r}^2+{a}^2}{{\Delta}}\left({r}^2+{a}^2-{a}\lambda\right)-{a}\left({a} \sin^2\theta-\lambda\right) \, , \label{t_eq_photon}
\end{align}
where
\begin{align}
&\lambda\equiv\frac{L}{E} \quad {\rm and} \quad \eta\equiv\frac{C}{E^2} .
\end{align}
The symbols $\pm_r={\rm sign}(u^{r})$ and $\pm_{\theta}={\rm sign}(u^{\theta})$ are defined by the 4-velocity of the light.
Moreover, the radial and angular potentials $R({r})$ and $\Theta(\theta)$ are, respectively, obtained as
\begin{align}
&R({r})=\left({r}^2+{a}^2-{a}\lambda \right)^2-{\Delta}\left[\eta +\left({a}-\lambda\right)^2 \right]\, ,\label{Rpotential}\\
&\Theta(\theta)=\eta+{a}^2\cos^2\theta-\lambda^2\cot^2\theta \, .
\label{Thetapotential}
\end{align}
All equations can be fully decoupled in terms of the Mino time $\tau$ becoming
\be
\frac{dx^{\mu}}{d\tau}\equiv\frac{\Sigma}{E}\frac{dx^{\mu}}{d\sigma}\, .
\label{tau'}
\ee
For the source point $x_{i}^{\mu}$ and the observer point $x^{\mu}$, the integral forms of the solutions now become \cite{gralla-2020, wang-2022}
\ba
\tau-\tau_{i} &=&I_{r}=G_{\theta} \, ,\label{r_theta}\\
\phi-\phi_{i} &=& I_{\phi}+{\lambda} G_{\phi}\, , \label{phi}\\
t-t_{i} &=& I_{t}+a^2G_{t} \, ,\label{t}
\ea
where
\ba
I_{r} &\equiv& \int_{r_{i}}^{r}\frac{1}{\pm_r\sqrt{R(r)}}dr,\quad
G_{\theta} \equiv \int_{\theta_{i}}^{\theta}\frac{1}{\pm_{\theta}\sqrt{\Theta(\theta)}}d\theta \, ,\label{Ir}\\
I_{\phi} &\equiv& \int_{r_{i}}^{r}\frac{{a \left[\left(2Mr-M^2+a^2\right)-a\lambda\right]}}{\pm_r\Delta\sqrt{R(r)}}dr,\quad
G_{\phi} \equiv \int_{\theta_{i}}^{\theta}\frac{\csc^2\theta}{\pm_{\theta}\sqrt{\Theta(\theta)}}d\theta \, ,\label{Iphi}\\
I_{t} &\equiv& \int_{r_{i}}^{r}\frac{{r^2\Delta+(2Mr-M^2+a^2)\left[\left(r^2+a^2\right)-a{\lambda}\right]}}{\pm_r\Delta\sqrt{R(r)}}dr,\quad
G_{t}\equiv\int_{\theta_{i}}^{\theta}\frac{\cos^2\theta}{\pm_{\theta}\sqrt{\Theta(\theta)}}d\theta \, . \label{It}
\ea
Notice that the radial potential $R(r)$ in (\ref{Rpotential}) can be rewritten as a quartic polynomial.
There are four roots, namely,
\begin{equation} \label{R_root}
R(r)=(r-r_1)(r-r_2) (r-r_3) (r-r_4),
\end{equation}
where $r_{1}< r_{2}\le r_{3} \le r_{4}$ in the cases that all of them are real valued with the property $r_{1}+r_{2}+r_{3}+r_{4}=0$ (see the detailed solutions of the roots in \cite{wang-2022} and Appendix \ref{appB}).
So, in the next section, we will classify all the roots and find the parameter space of the light rays for different types of the orbits.
Then the analytical solutions of the orbits will be solved.
Additionally, the analytical solutions may shed light on the light deflection by the black holes from the sources to the observers (for the schematic diagram, see Fig. 1 in \cite{hsieh-2021A, hsieh-2021B}), where the deflection angle $\hat{\alpha}$ can be obtained from solving (\ref{r_eq_photon}) and (\ref{phi_eq_photon}) at $\theta=\pi/2$ as
\be \label{deflection_angle}
\hat{\alpha} +\pi = 2 \int_{r_0}^{\infty} \frac{ 2Mr(a-\lambda) -(a-\lambda)(M^2-a^2) +r^2\lambda}{ \pm_r \Delta} \frac{1}{\sqrt{(r-r_1)(r-r_2) (r-r_3) (r-r_4)}} dr
\ee
with the turning point at $r_0=r_4$ to be examined later.
The value of $\lambda$ also called the impact parameter, as a function of the turning point $r_0$ is given by \cite{hsieh-2021A}
\be \label{br0_kn}
\lambda(r_0) =\frac{s \left( 2aM-a\frac{Q^2}{r_0} \right) -r_0 \sqrt{ \left( a\frac{Q^2}{r_0^2}-a\frac{2M}{r_0} \right)^2
+\left(1-\frac{2M}{r_0}+\frac{Q^2}{r_0^2}\right) \left[a^2(1+\frac{2M}{r_0}-\frac{Q^2}{r_0^2})+r_0^2 \right]}}
{ 2M-r_0-\frac{Q^2}{r_0}} \,
\ee
in the extremal limit $a^2+Q^2=M^2$, where $\lambda$ carries the sign with its absolute value denoted by
\begin{equation}
\varlambda = \vert \lambda \vert .
\end{equation}

\section{Roots of the radial potential and light rings of the extremal black holes}
\label{secIII}
\begin{figure}[h]
    \centering
    \includegraphics[width=16cm]{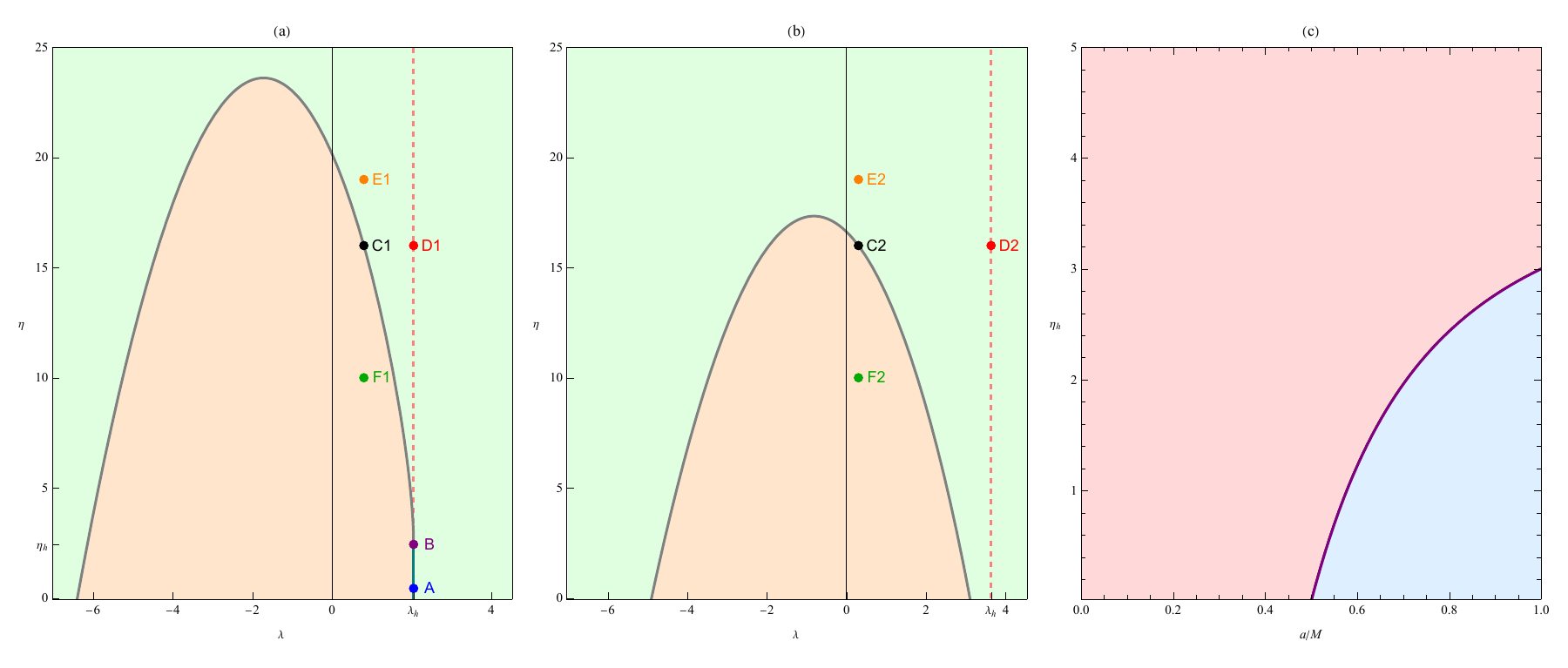}
    \caption{
    \baselineskip=17pt
    Diagram of parameter space $(\lambda, \eta)$ in (a) for $a =0.8M$ and (b) for $a=0.3M$, where the relevant roots of our study are three outermost roots $r_2$, $r_3$, and $r_4$ in this illustrative example.
    (a) for $a\ge M/2$, the red dashed line with a fixed value of $\lambda=\lambda_h$ and $\eta >\eta_h$ represents the parameters D1 with the stable double root $r_3=r_2=r_h$ at the horizon in Sec. \ref{StableDouble} while the gray solid line drawn from (\ref{tilde_lambda}) and (\ref{tilde_eta}) for $\eta >\eta_h$ represents the parameters C1 with the unstable double root $r_4=r_3 =r_u >r_h$ outside the horizon in Sec. \ref{UnstableDoubleU}.
    Two double roots merge and constitute the triple root at the point B with $\eta=\eta_h$ in Sec. \ref{TripleRoot}.
    Then the light blue line with a fixed value of $\lambda =\lambda_h$ and $\eta <\eta_h$ represents the parameters A with the unstable double root $r_4=r_3=r_h$ at the horizon in Sec. \ref{UnstableDoubleH}.
    The parameters E1 lie in the light green region with the four distinct roots in Sec. \ref{FourRoots} except for those on the red dashed line.
    The parameters F1 lie in the light orange region with two real-valued roots inside the horizon and a pair of the complex-valued roots in Appendix \ref{appC}.
    (b) for $a<M/2$, the red dashed line with a fixed value of $\lambda=\lambda_h$ represents the parameters D2 with the stable double root $r_3=r_2=r_h$ at the horizon in Sec. \ref{StableDouble}, while the gray solid line drawn from (\ref{tilde_lambda}) and (\ref{tilde_eta}) represents the parameters C2 with the unstable double root $r_u >r_h$ outside the horizon in Sec. \ref{UnstableDoubleU}.
    The parameters E2 lie in the light green region with the four distinct roots in Sec. \ref{FourRoots} except for them on the red dashed line.
    The parameters F2 lie in the light orange region with two real-valued roots inside the horizon and a pair of the complex-valued roots in Appendix \ref{appC}.
    (c) shows the value of $\eta_h$ as a function of the extremal black hole spin $a$ from (\ref{eta0}).
    The parameters lie in the light red region corresponding to the stable double root $r_3=r_2=r_h$ at the horizon; the parameters lie in the light blue region corresponding to the unstable double root $r_4=r_3=r_h$ at the horizon.}
    \label{eta_lambda_2}
\end{figure}

The spherical motion requires that the radial velocity and acceleration are zero, where $R(r_{\rm ss})=0$ and $R'(r_{\rm ss})=0$ with respect to the radius $r_{\rm ss}$, giving the equations for the double root as
\be
\left(r^2_{\rm ss} +a^2 -a\lambda\right)^2 -(r_{\rm ss} -M)^2\left[\eta+\left(\lambda-a\right)^2\right]=0\, ,
\label{R=0}
\ee
\be
2r_{\rm ss} \left(r^2_{\rm ss} +a^2 -a\lambda \right) -\left(r_{\rm ss} -M\right) \left[\eta+\left(\lambda-a\right)^2\right]=0\, .
\label{R'=0}
\ee
According to \cite{wang-2022}, the solutions of the parameters $\lambda$ and $\eta$ for the direct/retrograde motion are found to be
\begin{align}
\lambda_{\rm ss}&=a+\left[\frac{2M-r_{\rm ss}}{a}\right] r_{\rm ss}\, , \label{tilde_lambda}\\
\eta_{\rm ss}&=\left[\frac{4\left(Mr_{\rm ss}-M^2+a^2\right)-r_{\rm ss}^2}{a^2}\right]{r_{\rm ss}^2}\,. \label{tilde_eta}
\end{align}
Note that the radius of the spherical motion $r_{\rm ss}$ can have two types shown in Fig. \ref{eta_lambda_2}: one is outside the horizon, $r_u >M$ with the parameters along the gray line; the other is at the horizon $r_h =M$ with the parameters along the red dashed and light blue lines.
Their stability is determined by $R''(r)$,
\be
R''(r)=8r^2+4\left(r^2+a^2-a\lambda\right)-2\left[\eta+\left(\lambda-a\right)^2\right]\label{R'' equ}\, .
\ee
Substituting (\ref{tilde_lambda}) and (\ref{tilde_eta}) of the double root solutions gives
\be
R''(r_{\rm ss}) =8r_{\rm ss} \left(r_{\rm ss} -M\right)> 0
\label{R''whenr>M}
\ee
with $r_{\rm ss} =r_u >M$ showing again the known fact that those orbits are unstable.
This type of orbit occurs not only in extremal black holes but also in nonextremal ones \cite{wang-2022}.
As will be shown, they will contribute largely to the light ring to be studied later.

%
%

Nevertheless, there exists another type of the double root of $r_3=r_2$, which is stable for $R''(r_3=r_2)<0$.
This double root is inside the horizon for the nonextremal black holes but at the horizon for the extremal black holes.
For the double root $r_3=r_2=r_h$ at the horizon, from (\ref{R=0}), we can determine the value of $\lambda_h$ given by
\be
\lambda_h =\frac{a^2+M^2}{a} \, .
\label{l}
\ee
With $\lambda_h$, (\ref{R'=0}) is satisfied for all $\eta \ge 0$.
In the extremal Kerr case, the value of $\lambda_h$ reduces to $\lambda_h=2M$ in \cite{gralla-2018}.
The value of $\lambda_h$ decreases as $a$ increases due to the frame dragging effects.
The values of $\eta$ and $\lambda$ for the respective double roots are drawn in Fig. \ref{eta_lambda_2}.
%
\begin{figure}[h]
    \centering
    \includegraphics[width=0.8\textwidth]{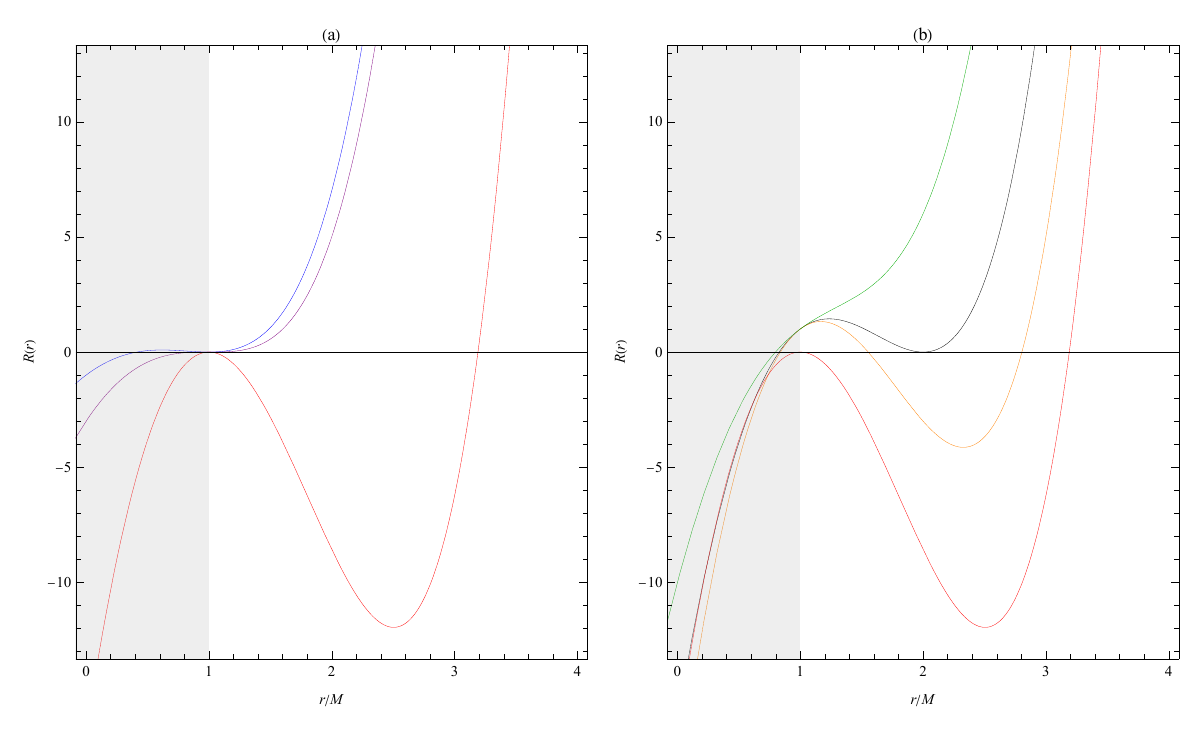}
    \caption{
    The graphics of the radial potential $R(r)$ versus the radius $r$ for $a=0.8M >M/2$ with the parameters in Fig. \ref{eta_lambda_2}:
    (a) the parameters along $\lambda =\lambda_h$ at A (blue), B (purple), and D1 (red);
    (b) the parameters at C1 (black), D1 (red), E1 (orange), and F1 (green).
    The similar plots for $a<M/2$ at C2, D2, E2, and F2 in Fig. \ref{eta_lambda_2} will be seen as in $a\ge M/2$ at their counterpart of C1, D1, E1, and F1 in (a).
    Shaded area indicates the inside of the horizon. }
    \label{radial potential}
\end{figure}
%

The instabilities of the double roots at the horizon can be studied with the value of $\lambda_h$ in (\ref{l}) by considering $R''(r)$ in (\ref{R'' equ}) given by
\be
R''(r) = 4 \left(3r^2-M^2 \right) -2\left(\eta+\frac{M^4}{a^2}\right)\, .
\label{R''when r=M}
\ee
For $\eta \ge 0$, $R''(r_h)=0$ can occur and imply that there exists a triple root with $r_4=r_3=r_2=r_h$ in the cases of $ a\geq M/2$, giving the value of $\eta_h$ to be
\begin{equation} \label{eta0}
\eta_h =M^2\left(4-\frac{M^2}{a^2}\right)\, .
\end{equation}
For $ a\ge M/2$, the triple root is pointed out at point B in Fig. \ref{eta_lambda_2}.
For $\eta >\eta_h$, the double root of $r_3=r_2=r_h$ is stable.
As $\eta$ decreases to the value of $\eta_h$, the stable double root ($r_3=r_2=r_h$) at the horizon and the unstable double root ($r_4=r_3=r_u >r_h$) will merge at the triple root with the parameters $\lambda_h$ and $\eta_h$ at the horizon.
Thus, with a fixed value of $\lambda_h$ and for $\eta<\eta_h$, the double root ($r_4=r_3=r_h$) at the horizon becomes unstable.
As for $ a<M/2$, $R''(r_h)<0$ for all $\eta >0$, where the double roots at the horizon are stable.
Thus, apart from the unstable double root of $r_u > r_h$, the radial potential has the isolated stable double root of $r_3=r_2=r_h$ at the horizon with the parameters shown in Fig. \ref{eta_lambda_2}.

The radial potentials $R(r)$ are drawn in Fig. \ref{radial potential} for $a\ge M/2$.
With the parameters along the line of $\lambda=\lambda_h$ in Fig. \ref{eta_lambda_2}, one can see from the radial potentials how the orbits with the radius $r_h$ are expected to change from the unstable one to the stable one by increasing the value of $\eta$ through the triple root at $\eta=\eta_h$.
The stable spherical orbits given by the stable double root are trapped with the radius of the double root $r_3=r_2=r_h$, while the unstable ones, when the light rays start from $ r_i \ge r_4=r_3=r_h$, can escape to spatial infinity.
Together with the emitted light rays with the parameters for $\eta \ge \eta_h$, giving the double root for the unstable spherical orbits ($r_u>r_h$), where the radial potentials are also depicted in Fig. \ref{radial potential}, they form the light ring in a D shape to be seen later.
For $a<M/2$, the stable double root $r_3=r_2=r_h$ occurs for all $\eta\ge 0$, leading to the stable spherical orbits trapped in the radius of the root $r_h$.
Nevertheless, the light ring is formed due to the escape of the light from near the double root $r_i>r_u>r_h$ of the unstable spherical orbits for all $\eta \ge0$.
The values of $\eta$ as a function of the black hole's spin $a$ for stable/unstable orbits are summarized in Fig. \ref{eta_lambda_2}.
We reproduce the results in \cite{ulbricht-2015, dolan-2016, khoo-2016}, and \cite{ogasawara-2020}, which analyze the parameter space of $\lambda$ and $\eta$ to render the unstable orbits and study the escape of the light emitted near the unstable orbits.

\begin{figure}[h]
    \centering
    \includegraphics[width=1.0\textwidth]{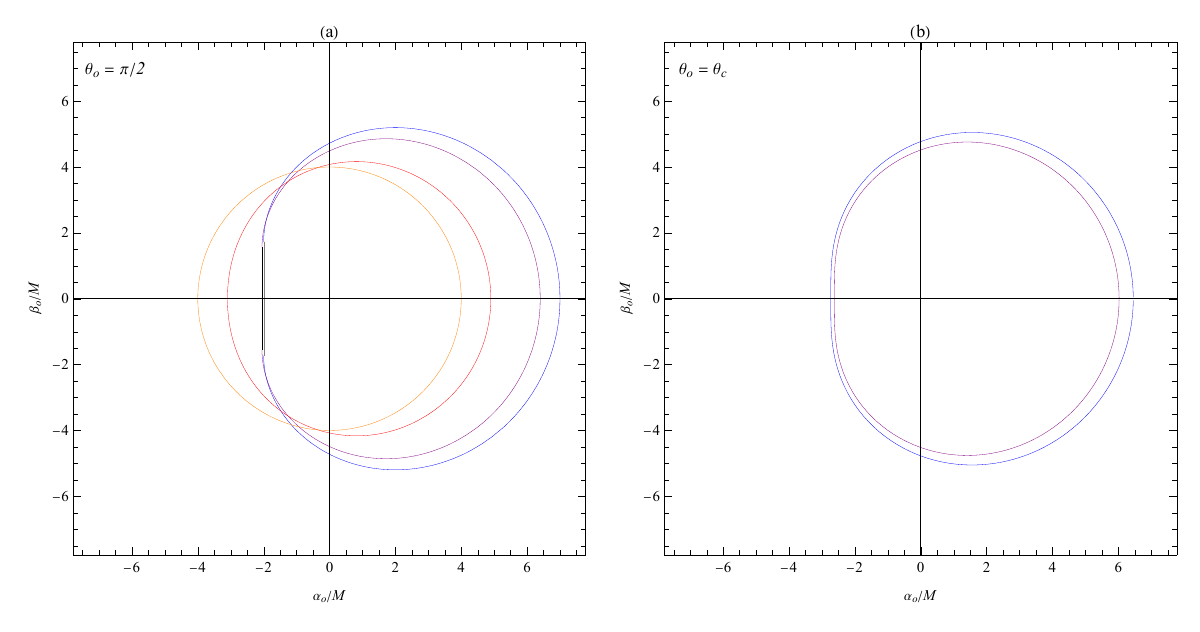}
    \caption{
    The light rings of the extremal black holes shown in the screen coordinates:
    (a) for $a=M$ (blue), $a=0.8M$ (purple), $a=0.3 M$ (red), and $a=0M$ (orange) at $\theta_o=\pi/2$;
    (b) for $a=M$ (blue) and $a=0.8M$ (purple), but at $\theta_o$ to be the corresponding critical values $\theta_{c}$ in (\ref{theta_c}).}
    \label{shadow}
\end{figure}

The ring of light appears at the edge of the black hole shadow to be seen on the screen coordinate $(\alpha_o ,\beta_o )$ in \cite{perlick-2021} of a two-dimensional plane far away from the black hole.
The light travels in the spherical orbit with a fixed ${r_{{\rm ss}}}$ at the double root, which can be obtained by rewriting (\ref{tilde_eta}) as
\begin{align} \label{rc_eta}
r_{\rm ss} \left[2 (M^2-a^2) + r_{\rm ss}^2- 3 M r_{\rm ss}\right] + s a \sqrt{D}=0 \;,
\end{align}
where
\begin{align} \label{D}
D =4r_{\rm ss}^2 \left[M r_{\rm ss}-(M^2-a^2) \right] -\left( {M-r_{\rm ss}}\right)^2 \eta_{\rm ss}
\end{align}
with $s=+\;(-)$.
Notice that the sign of $+$ ($-$) does not necessarily mean the direct (retrograde) orbit, which is determined by the sign of the corresponding $\lambda_{\rm ss}$.
In the special case of ${\eta=0}$ when the light travels along the equatorial circular orbits with a fixed ${r_{sc}}$, (\ref{rc_eta}) reduces to the one in \cite{hsiao-2020} but in the limit of the extremal black holes as
\begin{align} \label{rc_eq}
 2 (M^2-a^2) + r_{sc}^2- 3 M r_{\rm sc} + 2 s a \sqrt{M r_{\rm sc}-(M^2-a^2)}=0 \, .
\end{align}
The solution can be adapted from \cite{hsiao-2020} into
\begin{align}
r_{sc} & =\frac{3M}{2}
+\frac{1}{2\sqrt{3}}\sqrt{M^2+8a^2+U_{c}+\frac{P_{c}}{U_{c}}} \no\\
&\quad -\frac{s}{2\sqrt{3}}\sqrt{2M^2+16a^2-\left(U_c+\frac{P_{c}}{U_{c}}\right)
+\frac{24\sqrt{3}Ma^2}{\sqrt{M^2+8a^2+U_c+\frac{P_c}{U_c}}}} \;\; ,
\label{rc}
\end{align}
where
\begin{align}
P_{c} & =(M^2+8a^2)^2-24a^2(M^2+2a^2) \, , \\
U_{c} & =\bigg\{(M^2+8a^2)^3-36a^2(M^2+8a^2)(M^2+2a^2)+216M^2a^4 \bigg\}^\frac{1}{3} \, .
\end{align}
In the case of the extremal Kerr black holes $ a= M$, $r_{-c}= 4M$, and $r_{+c}=M$.
In addition,  combining (\ref{tilde_lambda}) and (\ref{rc_eta}) one can derive the following useful relation:
%
\begin{align}
\lambda_{\rm ss} & =a+s\frac{ 2 r_{\rm ss}^3-\left(r_{\rm ss}-M\right) \eta_{\rm ss}}{2\sqrt{r_{\rm ss}^2\left(Mr_{\rm ss}-M^2+a^2\right)-\left(\frac{r_{\rm ss}-M}{2}\right)^2 \eta_{\rm ss}}} \, .
\end{align}
When $\eta_{\rm ss}=0$, it also reduces to the known formula of the light orbits on the equatorial plane
\begin{align} \label{lambdasc}
\lambda_{sc} & =a+{s}\frac{r_{sc}^2}{\sqrt{Mr_{sc}-M^2+a^2}} \, .
\end{align}
The advantage of the above expression is to directly give the relation between the radius and the azimuthal angular momentum of the inner- and outermost unstable orbits of light.
For $a<M/2$, one can determine $r_{+c}$ and $r_{-c}$, respectively, from (\ref{rc}) and the associated $\lambda_{+c}$ and $\lambda_{-c}$ from (\ref{lambdasc}).
Through (\ref{tilde_lambda}) and (\ref{tilde_eta}), the values of $\eta$ and $\lambda$ for the unstable orbits can be obtained starting from the radius of $r_{-c}$ to $r_{+c}$ as the input to draw the light ring using (\ref{alpha_o}) and (\ref{beta_o}) to be introduced later in Fig. \ref{shadow}.
For $a \ge M/2$, the radius of the unstable spherical orbits is from ${r}_{-c}$ at $\eta=0$ in (\ref{rc}) to $r_{+c}=M$ at $\eta=\eta_h$.
Again, the values of $\lambda$ and $\eta$ can be obtained by (\ref{tilde_lambda}) and (\ref{tilde_eta}).
Then, while decreasing the value of $\eta$ from $\eta=\eta_h$ back to $\eta=0$, the values of $\lambda$ and $r_{+c}$ keep the constants as $\lambda=\lambda_h$ and $r_{+c}=M$, giving the known D shape light ring seen in Fig. \ref{shadow}.

The edge of the shadow is set by the threshold between captured and escaping light rays, which corresponds to unstable orbits \cite{de-vries-1999}.
The screen coordinates $(\alpha_o,\beta_o)$ of the observers are Cartesian coordinates for the position on the sky given by
\begin{align}
\alpha_o &=-\frac{\lambda}{\sin \theta_o}\, ,\label{alpha_o}\\
\beta_o &=\pm \sqrt{\eta + a^2 \cos^2 \theta_o - \lambda^2 \cot^2 \theta_o}\, ,\label{beta_o}
\end{align}
where $\theta_o$ is the polar angle of the observer \cite{Bardeen:1973tla}.
On the equatorial plane $\theta_0=\pi/2$, the light rings of extremal black holes for various $a$ are shown in Fig. \ref{shadow}.
For $a\ge M/2$, the D shape light ring is due to the existence of the unstable spherical orbits at the horizon contributing to the NHEKN line \cite{gralla-2018} with the end points $(\alpha_o, \beta_o)\vert_{r_{\rm ss}=r_h}$ obtained from $\eta=\eta_h$ and $\lambda=\lambda_h$.
As the observer's polar angle increases, two end points will be closed and the NHEKN line will gradually become invisible for the observer at $\theta_o < \theta_{c}$ with $\theta_{ c}$ by requiring $\beta_o \vert_{r_{\rm ss}=r_h} =0$ to be
\be \label{theta_c}
\theta_{c}=\cos^{-1}\left(\sqrt{\frac{-3 M^2 + 2M \sqrt{2 M^2 +a^2}}{a^2}}\right)\,.
\ee
For extremal Kerr black holes $a=M$, the above expression reduces to the known result in \cite{gralla-2018}.
Figure \ref{shadow} also shows the light ring at $\theta_o =\theta_{c}$, where the observer is at the critical angle.
%

\section{Analytic solutions for the escape/infall of light rays from/to black holes}
\label{secIV}
In this section, based on \cite{gralla-2020, wang-2022} we will provide the analytical expressions of light rays, which can escape/infall from/to black holes.
For $a\ge M/2$, their parameters of $\lambda$ and $\eta$ can be on the line of $\lambda_h$ shown in Fig. \ref{eta_lambda_2}, say at A of the double root for the unstable orbits $r_4=r_3=r_h$ and at B of the triple root $r_4=r_3=r_2=r_h$ as long as $r_i>r_h$.
Those light rays can either reach the horizon $r_h$ or escape to spatial infinity.
For completeness, let us also consider the light to be seen by observer in spatial infinity, which comes from the turning point $r_4$ outside the horizon, the largest real-valued root of the radial potential.
They can have the parameters at C1 or C2 (the unstable double root $r_4=r_3=r_u$), at D1 or D2 (the stable double root $r_3=r_2=r_h$), and at E1 or E2 (four different real-valued roots $r_4>r_3>r_2>r_1$) for $r_i >r_4$.
The more involved analytical solutions, plunging directly into the horizon or escaping into spatial infinity, are with the parameters at F1 or F2, where there are two real-valued roots of $r_1<0$ and $r_2<r_h$ and a pair of the complex-conjugate roots $r_3=r_4^*$ in \cite{ko-2024} to be summarized in Appendix \ref{appC}.
For the integrals of the radial potential $R(r)$ defined in Appendix \ref{appB}, in the cases of A, B, D1, and D2, where $\lambda$ restricts to (\ref{l}), $I_\phi (\tau)$ and $I_t(\tau)$ in (\ref{I phi by 1 2}) and (\ref{I t by 1 2}) can be obtained by taking $r_{\pm} \rightarrow r_h$ to be simplified as
\begin{align}
I_\phi (\tau) &= 2Ma I_{h}(\tau) \, ,\label{I phi app}\\
I_t(\tau) &= 2M \left({M^2+a^2}\right) I_h(\tau) +\left(3M^2+a^2\right) \tau +2M I_1(\tau) +I_2(\tau) \, \label{I t app}
\end{align}
with $I_h(\tau)$ defined in (\ref{I+-g}).
However, in the cases of the parameters at C1, C2, E1, E2, F1, and F2 in Fig. \ref{eta_lambda_2} with $\lambda \neq \lambda_h$, special care is required by taking the extremal limit.
In the limits of $r_+=r_h+\epsilon$ and $r_-=r_h-\epsilon$, the integral $I_\pm$ in (\ref{I+-g}) can be expanded as
\be
I_\pm=I_h \pm \epsilon\delta I_h
\ee
defined in (\ref{I+-g}).
From (\ref{I phi by 1 2}) and (\ref{I t by 1 2}), together with the limit of $r_\pm=r_h\pm\epsilon$, the expressions of ${I_\phi}(\tau)$ and $I_t(\tau)$ are given by
\begin{align}
I_\phi (\tau)=& 2Ma \left(I_h-\frac{a \lambda-a^2-M^2}{2M}\delta I_h\right) \, ,\label{I_phi_4}\\
I_t (\tau)=& 2M \left[ \left(2M^2+2a^2-a\lambda\right) I_h + \left( \frac{3 M^2+a^2}{2M}\right)\left( M^2+a^2- a\lambda\right)\delta I_h \right]\notag\\
&+ 2M I_1+I_2+(3M^2+a^2) \tau \, , \label{I_t_4}
\end{align}
as $\epsilon \to 0$.
The corresponding solutions related to the angular potential $\Theta(\theta)$ are summarized in Appendix \ref{appA}.

\subsection{Unstable double root at the horizon for $a\ge M/2$}
\label{UnstableDoubleH}
\begin{figure}[h]
    \centering
    \includegraphics[width=17cm]{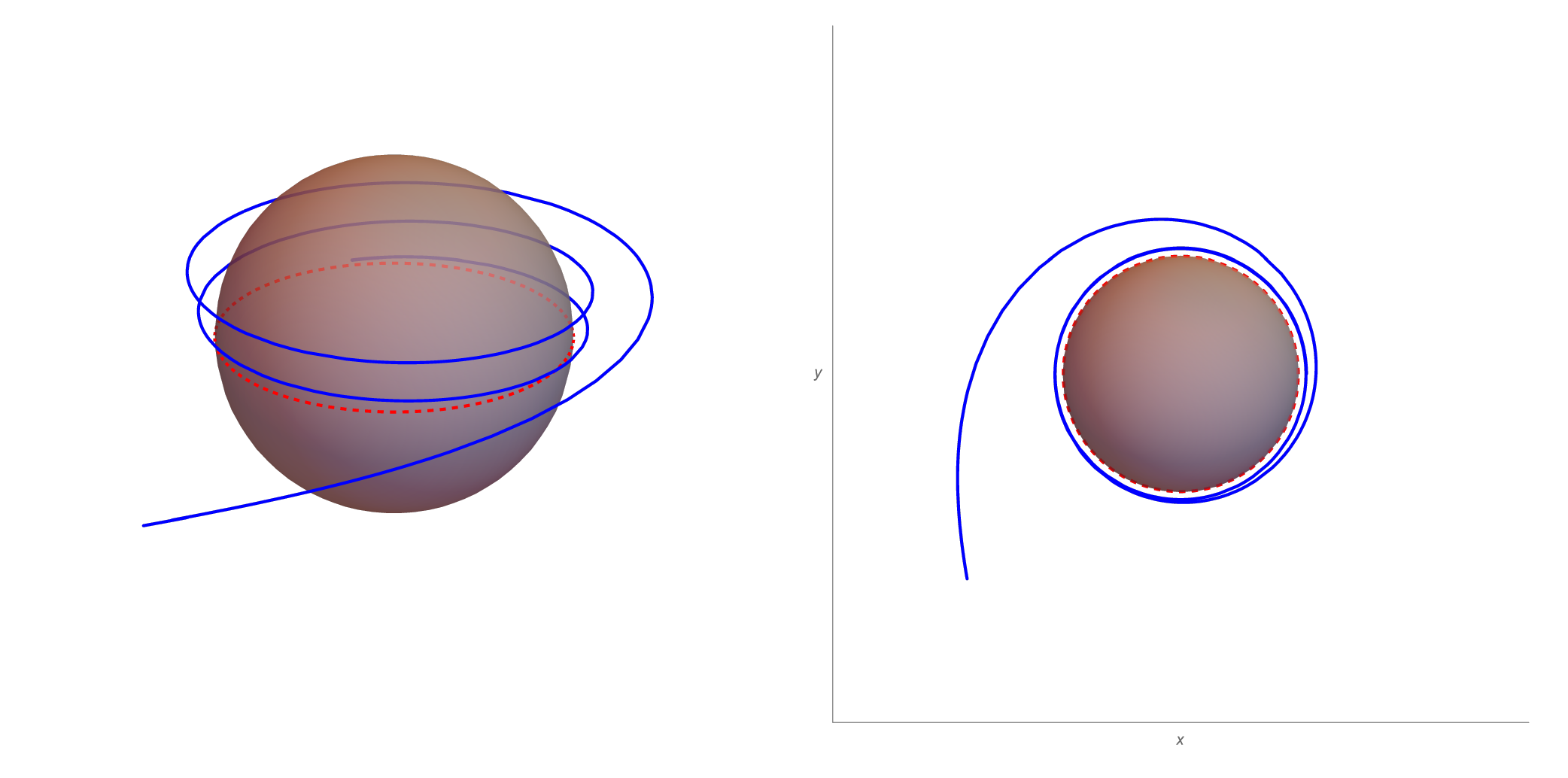}
    \caption{
    Illustration of a trajectory and its top view plots for $a=0.8M$ with the parameters at A in Fig. \ref{eta_lambda_2} for the unstable double root at the horizon.
    Similar plot will be seen with the parameters at B for the triple root at the horizon.
    {The shaded area indicates the inside of the horizon that fully overlaps with the light red shell showing the radius of the turning point. Finally, the dashed red circle lies on the equatorial plane.} }
    \label{r3=r43D}
\end{figure}

Let us start from the case of A in Fig. \ref{eta_lambda_2} with the parameters ($\lambda_h$, $\eta < \eta_h$) of the double root at the horizon $r_4=r_3=r_h$ for the unstable orbits.
For $r_i\ge r_h$, the orbits have been exclusively studied in \cite{porfyriadis-2017, gralla-2018}.
From the integrals (\ref{r_theta}) and (\ref{Ir}), the Mino time $\tau$ can be written as the function of $r$, 
\be
\tau^{U_h}(r) = -\frac{2 \nu_r}{\sqrt{\left(r_h-r_1\right)\left(r_h-r_2\right)}} \tanh^{-1}\sqrt{\frac{\left(r-r_1\right)\left(r_h-r_2\right)}{\left(r-r_2\right)\left(r_h-r_1\right)}} -\uptau^{U_h}_i
\label{tau of r unstable_h}
\ee
with $\nu_{r} =\pm_r
$, where the initial time $\uptau^{U_h}_i$ is chosen so that $\tau^{U_h}\left(r_i\right)=0$.
Its inversion leads to
\be
r^{U_h}(\tau) = \frac{r_1 \left(r_h-r_2\right) - r_2 \left(r_h-r_1\right) \tanh^2\left(X^{U_h}(\tau)\right)}{\left(r_h-r_2\right) -\left(r_h - r_1\right) \tanh^2\left(X^{U_h}(\tau)\right)} \, ,
\label{r tau unstable_h}
\ee
where
\be
X^{U_h}(\tau)=\frac{\sqrt{\left(r_h - r_1\right)\left(r_h - r_2\right)}}{2} \left(\tau^{U_h}+\uptau^{U_h}_i\right) \, .
\label{X_unstable_h}
\ee
We proceed by evaluating the coordinates $\phi(\tau)$ and $t(\tau)$ using (\ref{phi}) and (\ref{t}), which involve not only the angular integrals $G_{\phi}$ and $G_{t}$ in (\ref{G_phi_tau}) and (\ref{G_t_tau}), but also the radial integrals $I_{\phi}$ and $I_{t}$ in (\ref{Iphi}) and (\ref{It}) expressed in (\ref{I phi by 1 2}) and (\ref{I t by 1 2}).
In the extremal limits of $r_+$, $r_- \rightarrow r_h$ and the double root $r_4=r_3 \rightarrow r_h$, $I^{U_h}_{\pm}$ defined (\ref{I+-g}) reduces to $I^{U_h}_h$ and then (\ref{I phi by 1 2}) and (\ref{I t by 1 2}) to (\ref{I phi app}) and (\ref{I t app}), where
\begin{align}
&{I^{U_h}_{1}(\tau)=\frac{2r_h}{\sqrt{\left(r_h-r_2\right)\left(r_h-r_1\right)}}
X^{U_h}(\tau)-2\tanh^{-1}\left(\sqrt{\frac{r_h-r_2}{r_h-r_1}}
{\frac{1}{\tanh X^{U_h}(\tau)}}\right)-\nu_{r_i}\mathcal{I}^{U_h}_{1i}} \, ,\label{TUh1}\\
&{I^{U_h}_{2}(\tau)=\nu_r\sqrt{\left(r^{U_h}(\tau)-r_1\right)\left(r^{U_h}(\tau)-r_2\right)}+\frac{2 r_h^2}{\sqrt{\left(r_h-r_1\right)\left(r_h-r_2\right)}}X^{U_h}(\tau)-\nu_{r_i}\mathcal{I}^{U_h}_{2i}
}\, , \label{IUh2}\\
&{I^{U_h}_{h}(\tau)=\frac{1}{\sqrt{r_h-r_1}\left(r_h-r_2\right)^{3/2}} \frac{r^{U_h}(\tau)-r_2}{r^{U_h}(\tau)-r_h}\tanh X^{U_h}(\tau)-\frac{4r_h}
{\left(r_h-r_2\right)^{3/2} \left(r_h-r_1\right)^{3/2}} X^{U_h}(\tau)-\nu_{r_i}\mathcal{I}^{U_h}_{hi}}\, .
\label{IUhpm}
\end{align}
Notice that $\mathcal{I}_{hi}$, $\mathcal{I}_{1i}$, and $\mathcal{I}_{2i}$ are obtained by evaluating ${I}_{h}$, ${I}_{1}$, and ${I}_{2}$ at $r=r_i$ so that ${{I}_{h}^{U_h}(0)={I}_{1}^{U_h}(0)={I}_{2}^{U_h}(0)=0}$.
When $r_i=r_h +\epsilon$ near the horizon, $\uptau^{U_h}_i \propto \log \epsilon$ giving $X^{U_h} (0) \propto \log \epsilon$, which in turn leads to the divergence of $\mathcal{I}_{1i}^{U_h}$, $\mathcal{I}_{2i}^{U_h}$ $\propto$ $\log \epsilon$.
Apart from $\log \epsilon$ divergence, the leading divergent in $\mathcal{I}_{hi}^{U_h}$ is given by $1/{\epsilon}$.
They have been discussed in \cite{gralla-2018} for extremal Kerr black holes.
Alternatively, for the case of $r_i>r_h$, where the light travels toward black holes to reach near the horizon $r_h +\epsilon$, together with the finite value of $G_{\phi}$ and $G_{t}$, $\phi^{U_h}(r)$ and $t^{U_h}(r)$ have both $1/{\epsilon}$ and $\log \epsilon$ divergences.
Now we consider the equatorial motion by taking $\theta=\pi/2$ and $\eta\rightarrow0 $ limits in the results above.
In this case, $G_{\phi}=\tau$ and the solution of $\phi$ in (\ref{phi}) and $t$ in (\ref{t}) can be simplified to be
\begin{align}
\phi^{U_h}(r) =& \nu_r \left[\frac{-2Ma\sqrt{\left(r-r_1\right)\left(r-r_2\right)}}{\left(r-r_h\right)\left(r_h-r_1\right)\left(r_h-r_2\right)}\right. \notag \\
&+\left. \frac{8Ma r_h-2\lambda \left(r_h-r_2\right)\left(r_h-r_1\right)}{\left(r_h-r_2\right)^{3/2}\left(r_h-r_1\right)^{3/2}}\tanh^{-1}\sqrt{\frac{\left(r-r_1\right)\left(r_h-r_2\right)}{\left(r-r_2\right)\left(r_h-r_1\right)}}\right]\,
\end{align}
and
\begin{align}
t^{U_h}(r) =& \nu_r\left[\frac{2M\left(3M^2+a^2\right)\sqrt{\left(r-r_1\right)\left(r-r_2\right)}}{\left(r-r_h\right)\left(r_h-r_1\right)\left(r_h-r_2\right)}
+\sqrt{\left(r-r_1\right)\left(r-r_2\right)}
+4M\tanh^{-1}\sqrt{\frac{r-r_2}{r-r_1}} \right. \notag \\
&+ \left. \left(\frac{8M\left(3M^2+a^2\right)r_h}{\left(r_h-r_2\right)^{3/2}\left(r_h-r_1\right)^{3/2}}-\frac{8M^2-2(M^2-a^2)-4Mr_h-2r_h^2}{\sqrt{\left(r_h-r_1\right)\left(r_h-r_2\right)}}\right) \tanh^{-1}\sqrt{\frac{\left(r_1-r_1\right)\left(r_h-r_2\right)}{\left(r-r_2\right)\left(r_h-r_1\right)}} \right]\, .
\end{align}
In the limit of $a=M$ for extremal Kerr black holes, they reduce to the formulas in \cite{mummery-2023}.
The expressions $\phi^{U_h}(r)$ and $t^{U_h}(r)$ with $1/\epsilon$ and $\log{\epsilon}$ divergences as $r=r_h+\epsilon$ can also be seen in the above formulas.
Those orbits, which travel either from near the horizon to the spatial infinity or from the outside of the horizon toward the horizon, are of great interest to explore the nature of the horizon from the emitted light.
Figure \ref{r3=r43D} shows the illustration of the trajectory and its top view plots with the parameters at A for $a\ge M/2$.
%

The above divergence of $\phi^{U_h}(r)$ in the limit of $r=r_h+\epsilon$ on the equatorial plane has an interesting implication for the light bending effects due to the black holes.
As the turning point $r_0=r_4$ approaches the double root of $r_4=r_3=r_h$ by changing $\lambda$ along the line of $\eta=0$ toward $\lambda_h$ in (\ref{l}), namely, $r_0=r_h +\epsilon$, the deflection angle might also have the leading order divergence of a power law in terms of $1/\epsilon$, which is dramatically different from the logarithmic divergence for the corresponding double root in the nonextremal black holes \cite{hsieh-2021A}.
Following \cite{hsieh-2021A, hsieh-2021B, hsieh-2024}, one can examine the divergent part of the deflection angle in (\ref{deflection_angle}) as the unstable double root at $r_4=r_3=r_h$ is approached from the turning point through the change in $\varlambda$.
To do the integration in (\ref{deflection_angle}), we change the variable to $z$,
\be\label{z}
z \equiv 1-\frac{r_0}{r}
\ee
and the integral can be rewritten as
\be
\hat\alpha(r_0) = I (r_0) -\pi \, , \quad I(r_0)= \int_0^1 f(z,r_0) dz \,.
\ee
From \cite{hsieh-2021A} in the extremal case, the integrant giving the divergence of the integral is recalled as
\be \label{fD_double}
f_D(z,r_0)= \left\{ 1 +\frac{a^3 (z-1)^2-a^2 \varlambda (z-1)^2-a M (z-1) [M (z-1)+2 r_0] }{\varlambda [M (z-1)+r_0]^2} \right\} \frac{2r_0}{\sqrt{c_1(r_0) z +c_2(r_0) z^2}},
\ee
where
\be
\begin{split}\label{c12}
c_1(r_0)=&4(M^2-a^2)\left(1-\frac{a}{\varlambda}\right)^2-6Mr_0\left(1-\frac{a}{\varlambda}\right)^2+2r_0^2\left(1-\frac{a^2}{\varlambda^2}\right)\,,\\
c_2(r_0)=&-6(M^2-a^2)\left(1-\frac{a}{\varlambda}\right)^2+6Mr_0\left(1-\frac{a}{\varlambda}\right)^2-r_0^2\left(1-\frac{a^2}{\varlambda^2}\right)\, .
\end{split}
\ee
The coefficient $c_1(r_h) \to 0$ through (\ref{br0_kn}) in its extremal limit, but $c_2(r_h)$ is nonzero, which can also be realized from the expression of the radial potential in terms of the roots (\ref{R_root}) evaluated at the double root.
Thus, it is expected that the values of $z$ near the $z \to 0$ region will contribute significantly to the integral, resulting in the divergence.
If so, one can set a lower integration limit at $z\rightarrow c_1/c_2 \sim \epsilon/M$, leading to
\be
\begin{split}
I_D(r_0) \simeq & \int_0^1 \lim_{z\to 0} f_D(z,r_{h}) dz\\
       =& \int_{z \to \epsilon/M}  \left[\frac{4M a^2}{\sqrt{c_2(r_h)} (M^2+a^2)} \frac{1}{z^2}
       +\frac{2 M \left(M^2-a^2\right)}{\sqrt{c_2(r_h)} (M^2+a^2)} \frac{1}{z} \right] dz
       +{\rm finite \, part} \\
       =& -\frac{4M^2 a^2}{\sqrt{c_2(r_h)} (M^2+a^2)} \frac{1}{\epsilon}
       +\frac{2 M \left(M^2-a^2\right)}{\sqrt{c_2(r_h)} (M^2+a^2)} \log{\left(\frac{\epsilon}{M}\right)}
       +{\rm finite \, part}\, . \\
\end{split}
\ee
Therefore, the deflection angle in the SDL has such divergence behavior
\be \label{d_angle_dr_h}
\hat{\alpha}(r_0) \approx -\check{a} \left(\frac{r_0}{r_{h}}-1\right)^{-1} -\check{b} \log{ \left(\frac{r_0}{r_{h}}-1\right)} +\check{c},
\ee
where $\check{a}$ and $\check{b}$ can be read from above, while $\check{c}$ can be computed numerically.
This power-law divergence is a striking difference from the cases of the double root outside the horizon in nonextremal black holes \cite{hsieh-2021A}.
Thus, the analytical solutions of the light rays can not only trace the rays emitted from the horizon to the spatial infinity but also shed light on the light deflection by black holes from the sources to the observers.

\subsection{Triple root at the horizon for $a\ge M/2$}
\label{B}
\label{TripleRoot}
Another interesting orbit, which carries information of the horizon to the observer in the spatial infinity, is the case of the triple root of the radial potential $R(r)$, namely, $r_4=r_3=r_2=r_h>r_1$.
The initial condition $r_i>r_h$ is considered for light rays traveling toward or away from black holes with $(\lambda_h,\eta_h)$ in case B in Fig. \ref{eta_lambda_2}.
The solution along the $r$ direction can be obtained from (\ref{Ir}) and (\ref{r_theta}) as
\be
\tau^{T_h}(r)=\nu_r \frac{2}{r_1-r_h}\sqrt{\frac{r-r_1}{r-r_h}}-\uptau^{T_h}_i \, .
\label{tau of r triple}
\ee
Then the inverse of (\ref{tau of r triple}) leads to
\be
{r^{T_h}}(\tau)=\frac{r_1-r_h {X^{T_h}(\tau)}^2}{1-{X^{T_h}(\tau)}^2} ,
\label{r tau triple}
\ee
where
\be
{X^{T_h}(\tau)=\frac{r_1-r_h}{2} \left(\tau+\uptau^{T_h}_i\right)}\, .
\label{X tau of triple}
\ee
We then rewrite (\ref{I phi by 1 2}) and (\ref{I t by 1 2}) into (\ref{I phi app}) and (\ref{I t app}) with the integrals (\ref{I+-g}) and (\ref{Ing}) in the cases of a triple root given by
\begin{align}
&{I^{T_h}_1(\tau)=\frac{2 r_h}{\left(r_1-r_h\right)}X^{T_h}(\tau)+2\tanh^{-1}X^{T_h}(\tau)-\nu_{r_i}\mathcal{I}^{T_h}_{1i}}\, ,
\\
&{I^{T_h}_2(\tau)=\nu_r\sqrt{\left(r^{T_h}(\tau)-r_1\right)\left(r^{T_h}(\tau)-r_h\right)}-\frac{2 r_h^2}{r_h-r_1}X^{T_h}(\tau)-\nu_{r_i}\mathcal{I}^{T_h}_{2i}}\, ,\\
&I^{T_h}_{h}(\tau)= \frac{4}{3 (r_h-r_1)^2}\frac{r^{T_h}(\tau)+r_1}{r^{T_h}(\tau)-r_h}X^{T_h}(\tau)-\nu_{r_i}\mathcal{I}^{T_h}_{h i}\, .
\end{align}
For the initial coordinate $r$ near the triple root at the horizon
$r_i =r_h-\epsilon$,
$\uptau^{T_h}_i \propto 1/{\sqrt{\epsilon}}$, giving
$X^{T_h}(0) \propto 1/{\sqrt{\epsilon}}$ and in turn leading to
$\mathcal{I}_{1i}$, $\mathcal{I}_{2i} \propto 1/{\sqrt{\epsilon}}$, and
$\mathcal{I}_{hi} \propto 1/{\epsilon^{3/2}}$.
For the light from $r_i > r_h$ traveling toward the horizon and reaching $r=r_h+\epsilon$,
they take the coordinate time $t(r) \propto 1/{\epsilon^{3/2}}$ and have the change in azimuthal angle $\phi(r) \propto 1/{\epsilon^{3/2}}$.
Such orbits will take a longer coordinate time $t$ to arrive at spatial infinity as compared to the unstable double root.
The existence of the triple root of the radial potential for the null geodesics is unique for the extremal black holes.

In the case of $a= M/2$, giving $\eta=0$ on the equatorial plane, the orbits become simple as
\begin{align}
\phi^{T_h}(r) =& \nu_r \left[\frac{8}{3}\frac{Ma (r+r_1)}{\left(r-r_h\right)\left(r_h-r_1\right)^2}+\frac{2\lambda}{r_1-r_h}\right] \sqrt{\frac{r-r_1}{r-r_h}}\, ,\label{phi_T_h}
\end{align}
and
\begin{align}
t^{T_h}(r) =& \nu_r\left\{4M\tanh^{-1} \sqrt{ \frac{r-r_h}{r-r_1} } + \sqrt{ \left(r-r_1\right)\left(r-r_h\right)}\right. \notag \\
&+ \left. \frac{1}{r_1-r_h}\sqrt{\frac{r-r_1}{r-r_h}}\left[-\frac{8}{3}\frac{M\left(M^2+a^2\right)(r+r_1)}{\left(r-r_h\right)\left(r_h-r_1\right)}+2\left(3M^2+a^2\right)+4M r_h+2r_h^2\right]\right\} \label{t_T_h}\, .
\end{align}
Again, the behavior of $t^{T_h}(r) \propto 1/{\epsilon^{3/2}}$ and $\phi^{T_h}(r) \propto 1/{\epsilon^{3/2}}$ as $r=r_h+\epsilon$ can be seen in (\ref{phi_T_h}) and (\ref{t_T_h}).
The analytical expressions of the above two cases of the unstable double root and the triple root lying at the horizon $r_h$ for $a\ge M/2$ are the main results in this paper, which show how the light rays leave near the horizon and reach spatial infinity to be observed.
The solutions for both nonequatorial and equatorial motion above are remarkably simple in terms of elementary functions.
The trajectory plot with the parameters at point B is similar to that shown in Fig. \ref{r3=r43D}.

The above divergence of the azimuthal angle $\phi^{T_h}(r)$ will imply the significant power-law divergence of the deflection angle on the equatorial plane as $r_0$ approaches the triple root of $r_4=r_3=r_2=r_h$ at the horizon by changing $\varlambda$.
Let us consider the black hole spin $a= M/2$, where the triple root lies on the equatorial plane.
To incorporate the triple root effect, the integrant of the deflection angle in (\ref{deflection_angle}) giving the divergence will be extended further in \cite{hsieh-2021A} by including the $c_3(r_0) z^3$ term in (\ref{fD_double}) to be
\be
\begin{split}
f_D(z,r_0) =& \frac{2 \varlambda \left[3 M^2 (z-1)^2+8 M r_0 (z-1)+4 r_0^2\right] -M^2 (z-1) [3 M (z-1)+8 r_0]}{8 \varlambda [M (z-1)+r_0]^2} \\
&\times \frac{2r_0}{\sqrt{c_1(r_0) z +c_2(r_0) z^2 +c_3(r_0) z^3}} \label{fD_triple}
\end{split}
\ee
with an additional coefficient $c_3$ obtained as
\be
\begin{split}
c_3(r_0)=&4(M^2-a^2) \left(1-\frac{a}{\varlambda}\right)^2-2Mr_0\left(1-\frac{a}{\varlambda}\right)^2\, .
\end{split}
\ee
Through (\ref{br0_kn}) in the extremal limit, the coefficients $c_1(r_{h}) \to 0$, $c_2(r_{h}) \to 0$ defined in (\ref{c12}) and the coefficient $c_3(r_{h})$ is nonzero, which are expected from the radial potential in (\ref{R_root}) at the triplet root.
Again, setting the lower limit of the integral to account for the significant contributions from small $z$ as $z\rightarrow c_2/c_3 \sim \sqrt{c_1/c_3}\sim \epsilon/M$, the leading and subleading orders of the divergence can be found to be
\be
\begin{split}
I_D(r_0) \simeq & \int_0^1 \lim_{z\to 0} f_D(z,r_{h}) dz\\
    =& \int_{z \to \epsilon/M} \left[ \frac{4 M}{5 \sqrt{c_3(r_h)}} \frac{1}{z^{5/2}} +\frac{6 M}{5 \sqrt{c_3(r_h)} } \frac{1}{z^{3/2}}
    \right] dz +{\rm finite \, part}\\
\sim & -\frac{8 M^{5/2}}{15 \sqrt{c_3(r_{h})} } \frac{1}{\epsilon^{3/2}}
-\frac{12 M^{3/2}}{5 \sqrt{c_3(r_{h})} } \frac{1}{\epsilon^{1/2}}
+{\rm finite \, part} .
\end{split}
\ee
Therefore, the deflection angle in the SDL has a stronger divergence as expected shown as
\be
\hat{\alpha}(r_0) \approx -\check{a} \left(\frac{r_0}{r_{h}}-1\right)^{-3/2} -\check{b} \left(\frac{r_0}{r_{h}}-1\right)^{-1/2} +\check{c},
\ee
than the double root case in extremal black holes.
This provides an interesting observational effect in extremal black holes.

\subsection{Unstable double root outside the horizon}
\label{UnstableDoubleU}
In the case of the unstable double root $r_4=r_3=r_u>r_h$ at the points C1 and C2 with the parameters $(\lambda_{\rm ss}, \eta_{\rm ss})$ seen in Fig. \ref{eta_lambda_2} for both $a\ge M/2$ and $a<M/2$, when $r_i > r_4$, the light can escape to spatial infinity or travel toward the horizon until it settles at the double root $r_u$.
The expressions $\tau^U(r)$, $r^U(\tau)$, and $X^U(\tau)$ of the orbits follow the solutions in (\ref{tau of r unstable_h}), (\ref{r tau unstable_h}), and (\ref{X_unstable_h}) by replacing $r_h$ with $r_u$, respectively.
In addition, $I^U_1$ and $I^U_2$ have the formulas as in (\ref{TUh1}) and (\ref{IUh2}) again by replacing $r_h$ with $r_u$.
The expressions ${I_\phi}$ and $I_t$ also have the forms in (\ref{I_phi_4}) and (\ref{I_t_4}) where $I^U_1$, $I^U_2$ are given by
\vskip -0.2cm
\begin{align}
&I^{U}_{1}(\tau) = \frac{2r_u}{\sqrt{\left(r_u-r_2\right)\left(r_u-r_1\right)}}
X^U(\tau)-2\tanh^{-1}\left(\sqrt{\frac{r_u-r_2}{r_u-r_1}}
{\frac{1}{\tanh X^U(\tau)}}\right)-\nu_{r_i}\mathcal{I}^{U}_{1 i} ,\\
&I^{U}_{2}(\tau) = \nu_r\sqrt{\left(r^{U}(\tau)-r_1\right)\left(r^{U}(\tau)-r_2\right)}+\frac{2 r_u^2}{\sqrt{\left(r_u-r_1\right)\left(r_u-r_2\right)}}X^{U}(\tau)-\nu_{r_i}\mathcal{I}^{U}_{2i} .
\end{align}
However, the expression of $I^{U}_{h}$ becomes
\begin{align}
I^{U}_{h}(\tau) =& -\frac{2}{\left(r_u-r_h\right)\sqrt{\left(r_h-r_2\right)\left(r_h-r_1\right)}}
\tanh^{-1}{\left[\frac{\left(r_u-r_1\right)\left(r_h-r_2\right)}{\left(r_u-r_2\right)\left(r_h-r_1\right)}\tanh{X^U(\tau)}\right]} \notag \\
&+ \frac{2}{\left(r_u-r_h\right)\sqrt{\left(r_u-r_2\right)\left(r_u-r_1\right)}}X^U(\tau) -\nu_r\mathcal{I}^{U}_{hi}
\end{align}
and $\delta I^U_h$ is obtained as
\begin{align}
\delta I^U_h (\tau)
=& \frac{\nu_r \sqrt{(r(\tau) -r_1) (r(\tau) -r_2)}}{(r(\tau) -r_h) (r_h -r_1) (r_h -r_2) (r_u -r_h)}\notag \\
&+\frac{4 }{\sqrt{(r_h - r_1)(r_h - r_2)} (r_h - r_u)^2}
\tanh^{-1}{\left[\frac{\left(r_u-r_1\right)\left(r_h-r_2\right)}{\left(r_u-r_2\right)\left(r_h-r_1\right)}\tanh{X^U(\tau)} \right]}\notag \\
&-\frac{2\left[ (r_u+2r_h)r_u +r_1 r_2 \right] }{(r_h - r_1)^{3/2} (r_h - r_2)^{3/2} (r_h - r_u)^2}
\tanh^{-1}{\left[\frac{\left(r_u-r_1\right)\left(r_h-r_2\right)}{\left(r_u-r_2\right)\left(r_h-r_1\right)}\tanh{X^U(\tau)} \right]}\notag \\
&- \frac{2}{\sqrt{r_u - r_1} \sqrt{r_u - r_2} (r_h - r_u)^2}X^U(\tau) -\nu_r\mathcal{I}^{U}_{dhi}.
\label{delta Ih double roots}
\end{align}
The equatorial orbits then follow
\begin{align}\label{phi_U}
\phi^U(r) =& \nu_r\left\{-\frac{\sqrt{(r-r_1)(r-r_2)}}{(r-r_h)(r_h-r_1)(r_h-r_2)(r_u-r_h)}\left(-a^3+a^2\lambda -a M^2\right) \right.\notag \\
&+\left. \frac{4a M }{\sqrt{(r_h-r_1)(r_h-r_2)} (r_u-r_h)} \tanh^{-1}\sqrt{\frac{(r-r_1)(r_h-r_2)}{(r-r_2)(r_h-r_1)}} \right.\notag \\
&+\left. \frac{4 (a^3-a^2\lambda +aM^2) }{\sqrt{(r_h-r_1)(r_h-r_2)} (r_u-r_h)^2} \tanh^{-1}\sqrt{\frac{(r-r_1)(r_h-r_2)}{(r-r_2)(r_h-r_1)}} \right.\notag \\
&-\left. \frac{2 (a^3-a^2\lambda +aM^2) \left[ (r_u+2r_h)r_u +r_1 r_2 \right] }{(r_h-r_1)^{3/2}(r_h-r_2)^{3/2}(r_u-r_h)^2} \tanh^{-1}\sqrt{\frac{(r-r_1)(r_h-r_2)}{(r-r_2)(r_h-r_1)}} \right.\notag \\
&+ \left. \frac{2 \left[\left(-a^3+a^2\lambda -a M^2\right)-2a M(r_u-r_h)-\lambda(r_u-r_h)^2\right]}
{\sqrt{r_u-r_1}\sqrt{r_u-r_2}(r_u-r_h)^2} \tanh^{-1}\sqrt{\frac{(r-r_1)(r_u-r_2)}{(r-r_2)(r_u-r_1)}} \right\}\, ,
\end{align}
and
\begin{align}\label{t_U}
t^U(r) =& \nu_r\Bigg\{
\frac{\sqrt{(r - r_1)(r - r_2)}}{(r - r_h)(r_u - r_h)(r_h - r_1)(r_h - r_2)} \left[- 2M^2 a \lambda + (M^2+a^2)^2 + (M^2-a^2) a \lambda \right] \notag \\
&+\sqrt{(r - r_2)(r - r_1)}
+\frac{2 \left(-2 a \lambda M+4 M^2+4 M a^2\right) }{\sqrt{(r_h-r_1) (r_h-r_2)} (r_u-r_h)} \tanh^{-1}\sqrt{\frac{\left(r-r_1\right) \left(r_h-r_2\right)}{\left(r-r_2\right) \left(r_h-r_1\right)}}
\notag \\
&+4M \tanh^{-1}\sqrt{\frac{r -r_2}{r -r_1}}
-\frac{ 4[a \lambda M^2 +a^3 \lambda -(M^2+a^2)^2] }{\sqrt{(r_h-r_1) (r_h-r_2)} \left(r_u-r_h\right)^2} \tanh^{-1}\sqrt{\frac{\left(r-r_1\right) \left(r_h-r_2\right)}{\left(r-r_2\right) \left(r_h-r_1\right)}} \notag \\
&+\frac{ 2[a \lambda M^2 +a^3 \lambda -(M^2+a^2)^2] \left[ (r_u+2r_h)r_u +r_1 r_2 \right] }{(r_h-r_1)^{3/2} (r_h-r_2)^{3/2} (r_u-r_h)^2} \tanh^{-1}\sqrt{\frac{\left(r-r_1\right) \left(r_h-r_2\right)}{\left(r-r_2\right) \left(r_h-r_1\right)}} \notag \\
&-\frac{2 (2M r_u + r_u^2 + 3M^2 + a^2)}{\sqrt{(r_u - r_1)(r_u - r_2)}} \tanh^{-1}\sqrt{\frac{(r - r_1)(r_u - r_2)}{(r - r_2)(r_u - r_1)}} \notag \\
&-\frac{2(4M^3 - 2M a \lambda + 4M a^2) }{\sqrt{(r_u - r_1)(r_u - r_2)}(r_u - r_h)} \tanh^{-1}\sqrt{\frac{(r - r_1)(r_u - r_2)}{(r - r_2)(r_u - r_1)}} \notag \\
&-\frac{2 [- 2M^2 a \lambda + (M^2+a^2)^2 + (M^2-a^2) a \lambda]}{\sqrt{(r_u - r_1)(r_u - r_2)}(r_u - r_h)^2} \tanh^{-1}\sqrt{\frac{(r - r_1)(r_u - r_2)}{(r - r_2)(r_u - r_1)}} \Bigg\} \, .
\end{align}
%
Let us now study the anticipated divergence when $r_i=r_u +\epsilon$ is near the double root.
In this case, $\uptau^{U}_i \propto \log \epsilon$ giving $X^{U} (0) \propto \log \epsilon$, which in turn leads to the divergence of $\mathcal{I}_{1i}^{U}$, $\mathcal{I}_{2i}^{U}$ $\propto$ $\log \epsilon$.
Thus, the leading divergence in $\mathcal{I}_{hi}^{U_h}$ is given by $\log \epsilon$.
The coordinate time $t^{U}(r)$ and the azimuthal angle $\phi^{U}(r)$ then have the divergence of $\log \epsilon$, which can also be seen from the $\tanh^{-1}$ terms of the solutions on the equatorial plane (\ref{phi_U}) and (\ref{t_U}).
The behavior of the divergence is very different from the cases of the unstable roots at the horizon $r_h$ discussed in Sec. \ref{UnstableDoubleH}.
The typical trajectory plot has been shown in \cite{wang-2022}.

The divergence of $\phi^{U}(r)$ as the coordinate $r$ reaches the double root outside the horizon leads to the logarithmic divergence of the deflection angle as the double root is approached, namely,
\be
\hat{\alpha}(r_0) \approx -\check{b} \log{ \left(\frac{r_0}{r_{h}}-1\right)} +\check{c}
\ee
with $\check{a}=0$ in (\ref{d_angle_dr_h}).
Since the double root is outside the horizon, the deflection angle will find weaker divergence as compared with the cases of the double root at the horizon.
This has been shown in \cite{hsieh-2021A}.

\subsection{Stable double root at the horizon}
\label{StableDouble}
Another unique feature of extremal black holes is that there exist stable spherical orbits given by the double root of $r_3=r_2=r_h$ with the parameters $(\lambda_h, \eta >\eta_h )$ for $a\ge M/2$ and $(\lambda_h, \eta \ge 0 )$ for $a<M/2$ at the points D1 and D2 in Fig. \ref{eta_lambda_2}, respectively.
Since the light allows traveling for $R(r)>0$ with $r_i\geq r_4$, they can either approach the black holes up to $r=r_4$ and then bounce back or travel directly to the spatial infinity.
According to (\ref{Ir}) and (\ref{r_theta}), the orbits have the following solution:
\be
\tau(r) = \frac{2 \nu_r}{\sqrt{(r_h-r_1)(r_4-r_h)}} \tan^{-1}\sqrt{\frac{(r-r_4)(r_h-r_1)}{(r-r_1)(r_4-r_h)}} -\uptau^S_i
\label{tau of r stable}
\ee
and its inversion
\be
r^S(\tau)=\frac{r_4\left(r_h-r_1\right)-r_1\left(r_4-r_h\right)\tan^2\left(X^S(\tau)\right)}{\left(r_h-r_1\right)-\left(r_4-r_h\right)\tan^2\left(X^S(\tau)\right)},
\label{r tau stable}
\ee
where
\be
X^S(\tau)=\frac{1}{2} \sqrt{\left(r_h-r_1\right)\left(r_4-r_h\right)} \left(\tau^S +\uptau^S_i \right)\, .
\ee
With (\ref{I phi by 1 2}) and (\ref{I t by 1 2}) reducing to (\ref{I phi app}) and (\ref{I t app}), in the case of the stable double root, the integrals in (\ref{I+-g}) and (\ref{Ing}) become
\begin{align}
&{I^S_{1}(\tau)=\frac{2r_h}{\sqrt{\left(r_h-r_1\right)\left(r_4-r_h\right)}}X^S(\tau)+2 \tanh^{-1}\left(\sqrt{\frac{r_4-r_h}{r_h-r_1}}\tan X^S(\tau)\right)-\nu_{r_i}\mathcal{I}^S_{1i}} ,\\
&{I^S_{2}(\tau)=\nu_r\sqrt{\left(r^S(\tau)-r_4\right)\left(r^S(\tau)-r_1\right)}+\frac{2 r_h^2}{\sqrt{\left(r_h-r_1\right)\left(r_4-r_h\right)}}X^S(\tau)-\nu_{r_i}\mathcal{I}^S_{2i}} ,\\
&{I^S_{h }(\tau) = \frac{1}{\sqrt{r_4-r_h}\left(r_h-r_1\right)^{3/2} } \frac{r^S(\tau)-r_1}{r^S(\tau)-r_h} \tan{X^S(\tau)} +\frac{4 r_h}{\left(r_h-r_1\right)^{3/2} \left(r_4-r_h\right)^{3/2}}X^S(\tau)-\nu_{r_i}\mathcal{I}^S_{h i}} .
\end{align}
Although the orbits will not reach the stable double  root at the horizon,  the solutions certainly show the dependence of that root, from which one sees the difference from the orbits discussed in Sec. \ref{UnstableDoubleU}.
\begin{figure}[h]
    \centering
    \includegraphics[width=17cm]{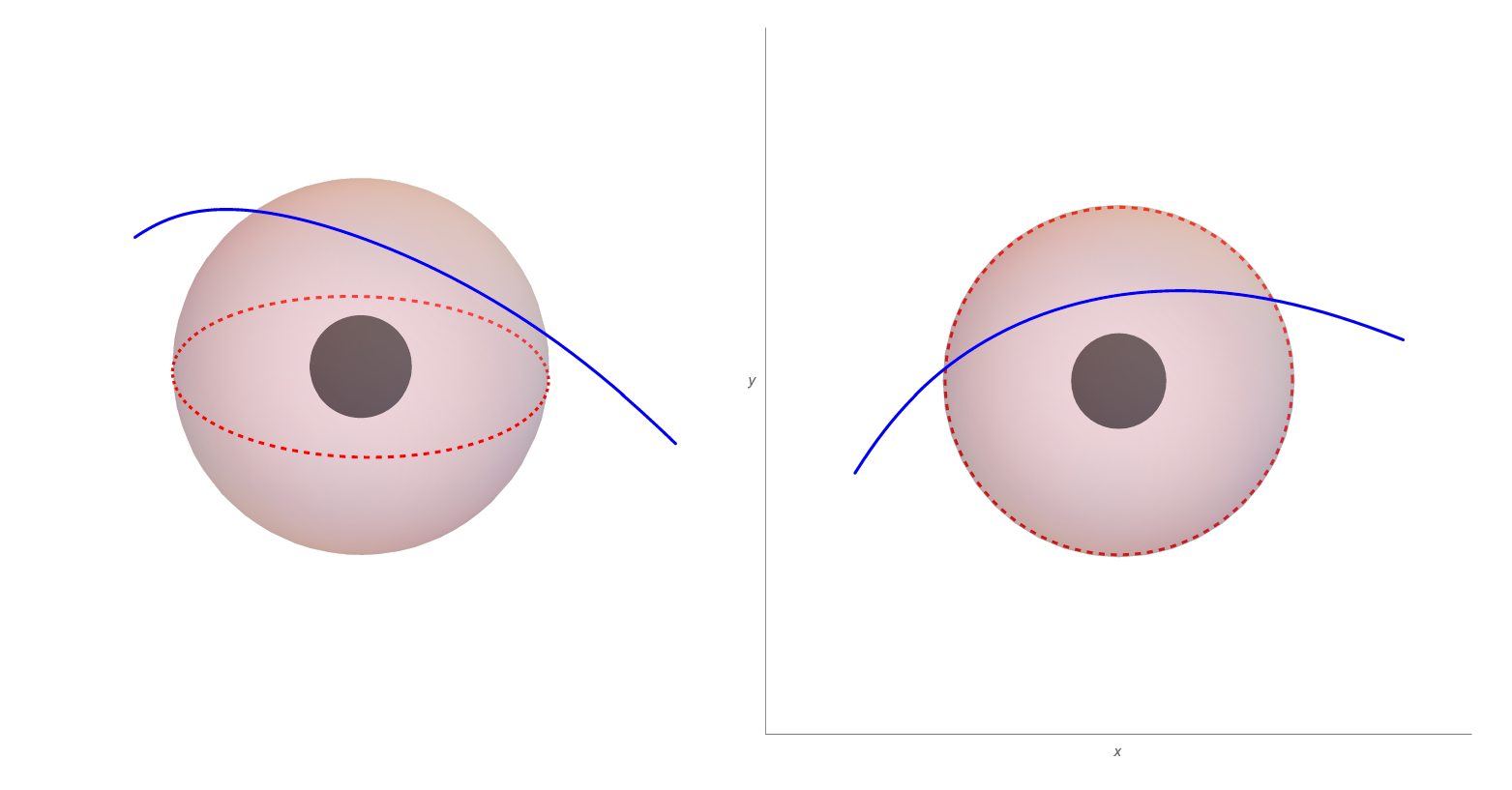}
    \caption{
    Illustration of a trajectory and its top view plots for $a=0.8M$ with the parameters at D1 in Fig. \ref{eta_lambda_2} for the stable double root at the horizon and a turning point at $r_0=r_4>r_h$.
    {The shaded area indicates the inside of the horizon and the light red shell represents the radius of the turning point. Finally, the dashed red circle lies on the equatorial plane.} }
    \label{r2=r33D}
\end{figure}
%

On equatorial orbits only for $a<M/2$ with $\eta=0$, we can rewrite the solution as
\begin{align}
&\phi^S(r) = \nu_r \left[\frac{2Ma\sqrt{\left(r-r_1\right)\left(r-r_4\right)}}{\left(r-r_h\right)\left(r_h-r_1\right)\left(r_4-r_h\right)}\right. \notag \\
&\quad\quad\quad\left. +\frac{8Ma r_h-2\lambda \left(r_h-r_1\right)\left(r_4-r_h\right)}{\left(r_h-r_1\right)^{3/2}\left(r_4-r_h\right)^{3/2}}\tan^{-1}\sqrt{\frac{\left(r-r_4\right)\left(r_h-r_1\right)}{\left(r-r_1\right)\left(r_4-r_h\right)}}\right] ,
\end{align}
and
\begin{align}
&t^S(r)=\nu_r\left\{\frac{2M\left(3M^2+a^2\right)\sqrt{\left(r-r_1\right)\left(r-r_4\right)}}{\left(r-r_h\right)\left(r_h-r_1\right)\left(r_4-r_h\right)}+4M\tanh^{-1}\sqrt{\frac{r-r_4}{r-r_1}} +\sqrt{\left(r-r_4\right)\left(r-r_1\right)}\right. \notag \\
&\quad\quad\quad\left.+\left[\frac{8M\left(3M^2+a^2\right)r_h}{\left(r_h-r_1\right)^{3/2}\left(r_4-r_h\right)^{3/2}}+\frac{6M^2+2a^2+4Mr_h+2r_h^2}{\sqrt{\left(r_h-r_1\right)\left(r_4-r_h\right)}}\right]
\tan^{-1}\sqrt{\frac{\left(r-r_4\right)\left(r_h-r_1\right)}{\left(r-r_1\right)\left(r_4-r_h\right)}} \right\} .
\end{align}
The cases of the stable double root at the horizon strongly simplify the solutions of the orbits.
Although the orbits will not reach the horizon, they can possibly escape from the black hole to the spatial infinity, where the trajectory plot is displayed in Fig. \ref{r2=r33D}.
The resulting light deflection from the sources to the observers has been studied in \cite{hsiao-2020} by taking the extremal limit.
The detailed deflection angle will be derived in the next section in a general azimuthal angular momentum $\varlambda$.
The corresponding deflection angle will be achieved in the limit of $\varlambda = \varlambda_h$.
Since the light rays will not reach the stable unstable double root but are deflected at the turning point, the light deflection by extremal black holes is not expected to be very different from that of nonextremal black holes.

\subsection{Four distinct real-valued roots}
\label{FourRoots}
Let us also consider the analytical solutions of the null geodesics with the parameters $\lambda$ and $\eta$ at the points E1 and E2 in Fig. \ref{eta_lambda_2} for both $a\ge M/2$ and $a<M/2$ although the orbits will not reach or come from the horizon.
In these cases, the radial potential has four distinct roots obtained by taking the extremal limit in \cite{wang-2022}.
The orbits start from $r_i > r_4$ and then they either escape from the black holes and reach spatial infinity or travel toward the black holes and bounce back at $r=r_4$ to spatial infinity.
For extremal black holes, the solutions can be achieved from \cite{wang-2022} by taking $r_{\pm} \rightarrow r_h$ to be
\be
r(\tau)=\frac{r_{4}(r_{3}-r_{1})-r_{3}(r_{4}-r_{1}){\rm sn}^2\left(X(\tau)\left|{k}\right)\right.}{(r_{3}-r_{1})-(r_{4}-r_{1}){\rm sn}^2\left(X(\tau)\left|{k}\right)\right.},
\label{r_tau_general}
\ee
where
\begin{align}
&X(\tau)=\frac{\sqrt{(r_{3}-r_{1})(r_{4}-r_{2})}}{2}\tau+\nu_{r_i} F\Bigg(\sin^{-1}\left(\sqrt{\frac{(r_{i}-r_{4})(r_{3}-r_{1})}{(r_{i}-r_{3})(r_{4}-r_{1})}}\right)\left|{k}\Bigg)\right.\,,\\
&k=\frac{(r_{3}-r_{2})(r_{4}-r_{1})}{(r_{3}-r_{1})(r_{4}-r_{2})}
\end{align}
where $\rm sn$ is the Jacobi elliptic sine function.
The expressions $I_\phi$ and $I_t$ in (\ref{I phi by 1 2}) and (\ref{I t by 1 2}) reduce to (\ref{I_phi_4}) and (\ref{I_t_4}) in the extremal limit where $I_{h}$, $I_1$, and $I_2$ are obtained as
\begin{align}
&I_{h}(\tau)=\frac{2}{\sqrt{(r_{3}-r_{1})(r_{4}-r_{2})}}
\left[\frac{X(\tau)}{r_{3}-r_{h}}+\frac{r_{3}-r_{4}}{(r_{3}-r_{h})(r_{4}-r_{h})}\Pi\left(\alpha_{h};\Upsilon_{\tau}\left|{k}\right)\right.\right]-\nu_r{\mathcal{I}_{hi}} \label{I_pm_tau}\;,\\
&I_{1}(\tau)=\frac{2}{\sqrt{(r_{3}-r_{1})(r_{4}-r_{2})}}\left[r_{3}X(\tau)+(r_{4}-r_{3})\Pi\left(\alpha;\Upsilon_{\tau}\left|{k}\right)\right.\right]-\nu_r{\mathcal{I}_{1i}} \label{I_1_tau}\;,\\
&I_{2}(\tau)=\nu_{r}\frac{\sqrt{\left(r(\tau)-r_{1}\right)\left(r(\tau)-r_{2}\right)\left(r(\tau)-r_{3}\right)\left(r(\tau)-r_{4}\right)}}{r(\tau)-r_{3}}-\frac{r_{1}\left(r_{4}-r_{3}\right)-r_{3}\left(r_{4}+r_{3}\right)}{\sqrt{(r_{3}-r_{1})(r_{4}-r_{2})}}X(\tau)\notag\\
&\;\;\quad\quad-\sqrt{(r_{3}-r_{1})(r_{4}-r_{2})}E\left(\Upsilon_{\tau}\left|{k}\right)\right.+\frac{\left(r_{4}-r_{3}\right)\left(r_{1}+r_{2}+r_{3}+r_{4}\right)}{\sqrt{(r_{3}-r_{1})(r_{4}-r_{2})}} \Pi\left(\alpha;\Upsilon_{\tau}\left|{k}\right)\right.-\nu_r{\mathcal{I}_{2i}} \label{I_2_tau}\;.
\end{align}
The parameters of the elliptical integrals given above are defined to be
\begin{align}\label{Upsilon}
&\Upsilon_{\tau}={\rm am}\left(X(\tau)\left|k\right)\right.=\nu_{r_i}\sin^{-1} \sqrt{\frac{\left(r(\tau)-r_{4}\right)(r_{3}-r_{1})}{\left(r(\tau)-r_{3}\right)(r_{4}-r_{1})}} \;,\\
&\alpha_{h}=\frac{(r_{3}-r_{h})(r_{4}-r_{1})}{(r_{4}-r_{h})(r_{3}-r_{1})} \;,
\hspace*{4mm}
\alpha=\frac{r_{4}-r_{1}}{r_{3}-r_{1}}\, \label{alpha}.
\end{align}
Moreover, $\delta I_h$ is computed as
\be
\delta I_h (\tau)
=\frac{2}{\sqrt{(r_4-r_2)(r_3-r_1)}} \big[ C_F X(\tau)
+ C_E E\left(\Upsilon_{\tau}|k \right)
+ C_\Pi \Pi\left(\alpha_{h};\Upsilon_{\tau}|k \right)
+ C_0(r(\tau)) \big]
- \nu_r\mathcal{I}_{dhi}
\label{delta Ih}
\ee
with the coefficients of the elliptical integrals
\begin{align}
& C_F = \frac{1}{2 (r_h - r_4) (r_h - r_1) (r_h - r_3)^2} \left[ 2 (r_h - r_1) (r_h - r_4) + (r_3 - r_1) (r_4 - r_3) \right] \notag ,\\
& C_E = -\frac{(r_3 - r_1)(r_4 - r_2)}{2(r_h - r_4)(r_h - r_1)(r_h - r_2)(r_h - r_3)} \notag ,\\
& C_{\Pi} = \frac{(r_4 - r_3)}{2(r_h - r_1)(r_h - r_3)^2(r_h - r_2)(r_h - r_4)^2} \left[ (r_3 r_4 - r_1 r_2)(r_3 + r_4 + 2r_h) + 2 r_h (2r_h^2 - r_1^2 - r_2^2) \right] \notag ,\\
&C_0(r) = -\frac{(r_4 - r_2)}{2(r_h - r_4)(r_h - r_1)(r_h - r_2)} \sqrt{\frac{r_3 - r_1}{r_4 - r_2}}\frac{\sqrt{(r-r_1)(r-r_2)(r-r_3)(r-r_4)}}{(r-r_h)(r-r_3)} \notag .
\end{align}
These solutions can be compared with those discussed in Secs. \ref{StableDouble} and \ref{FourRoots}, where all of them have the turning point at $r=r_4$ and will not reach the horizon.

On the equatorial plane, the solutions can be simplified as
\begin{align}
&\phi(r)= \frac{2\nu_r}{\sqrt{(r_3 - r_1)(r_4 - r_2)}} \Bigg\{ \left[ \frac{2Ma}{r_3 - r_h} - (a^2 \lambda - a^3 - aM^2) C_F + \lambda \right] F\left(\sin^{-1}(X(r))| k\right) \notag\\
&\quad\quad\quad
+ \left[ \frac{2Ma(r_3 - r_4)}{(r_3 - r_h)(r_4 - r_h)} - (a^2 \lambda - a^3 - aM^2) C_\Pi \right] \Pi\left(\alpha_h; \sin^{-1}(X(r))| k\right) \notag\\
&\quad\quad\quad
-(a^2 \lambda - a^3 - a M^2) C_E E\left(\sin^{-1}(X(r))|k\right) -(a^2 \lambda - a^3 - a M)C_0(r) \Bigg\}\, ,
\end{align}
and
\begin{align}
&t(r)= \frac{2 \nu_r}{\sqrt{(r_3 - r_1)(r_4 - r_2)}} \Bigg\{ \left[ \frac{(4M^3 - 2M a \lambda + 4M a^2)}{r_3 - r_h} - \frac{r_1 (r_4 - r_3) - r_3 (r_4 + r_3)}{2} \right] F\left(\sin^{-1}(X(r))| k\right) \notag\\
&\quad\quad\quad
+ \left[ (M^2+a^2) (M^2 +a^2 -a\lambda) C_F + 2M r_3 + 3M^2 + a^2 \right] F\left(\sin^{-1}(X(r))| k\right) \notag\\
&\quad\quad\quad
+ \left[ (M^2+a^2) (M^2 +a^2 -a\lambda) C_E - \frac{(r_3 - r_1)(r_4 - r_2)}{2} \right] E\left(\sin^{-1}(X(r))|k\right) \notag\\
&\quad\quad\quad
+ \frac{(r_3 - r_4) (4M^3 - 2M a \lambda +4M a^2) }{(r_3 - r_h)(r_4 - r_h)} \Pi\left(\alpha_h; \sin^{-1}(X(r))|k\right) \notag\\
&\quad\quad\quad
+ (M^2+a^2) (M^2 +a^2 -a\lambda) C_\Pi \Pi\left(\alpha_h; \sin^{-1}(X(r))|k\right) \Bigg\} \notag\\
&\quad\quad\quad
+\nu_r \left\{ \frac{4 M (r_4 - r_3)}{\sqrt{(r_3 - r_1)(r_4 - r_2)}} \Pi\left(\alpha; \sin^{-1}(X(r))|k\right)
+ \frac{\sqrt{(r -r_1)(r -r_2)(r -r_3)(r -r_4)}}{r -r_3}+C_0(r) \right\}\, .
\end{align}
The typical trajectory plot for $r_i > r_4$ can be seen in \cite{wang-2022}.
Also, the light rays from the sources meet the turning point and then
are deflected by the black hole to another direction, which has been studied in \cite{hsiao-2020} in the extremal limit.
Let us now consider the limit of the largest root $r_4$ of the radial potential to be very far away from the horizon, namely $r_0 \gg M$ or $\varlambda \gg M$; the deflection angle in \cite{hsiao-2020} can be written in an expansion of a small parameter $M/r_0$ to be
\be
\begin{split}
\hat{\alpha}(r_0) =& 4 \left(\frac{M}{r_0} \right)
+\left[ -4 (sa_{\bullet}+1) +\frac{3\pi}{4} \left(a_{\bullet}^2+4\right) \right] \left(\frac{M}{r_0} \right)^2\\
&+\left[ \frac{16}{3} (3a_{\bullet}^2 +3sa_{\bullet} +5) -\frac{\pi}{2} (2sa_{\bullet}^3 +3a_{\bullet}^2 +18sa_{\bullet} +12) \right] \left(\frac{M}{r_0} \right)^3
+\mathcal{O}\left(\frac{M}{r_0} \right)^4
\end{split}
\ee
with the unit spin $a_{\bullet} \equiv a/M$ in the so-called weak deflection limit (WDL).
Further, we can replace $r_0$ by $\varlambda$ through (\ref{br0_kn}) in the extremal limit and obtain
\be
\begin{split}
\hat{\alpha}(\varlambda) = 4 \left(\frac{M}{\varlambda} \right)
+\frac{3 \pi a_{\bullet}^2 -16 s a_{\bullet} +12 \pi}{4} \left(\frac{M}{\varlambda} \right)^2
+\frac{-3 \pi s a_{\bullet}^3 +60 a_{\bullet}^2 -27 \pi s a_{\bullet} +80}{3} \left(\frac{M}{\varlambda} \right)^3
+\mathcal{O}\left(\frac{M}{\varlambda} \right)^4.
\end{split}
\ee
Then, when the azimuthal angular momentum $\varlambda$ lies at $\varlambda_h$ in (\ref{l}), the deflection angle will be the case in Sec. \ref{StableDouble}.
Not much difference in the light deflection in the WDL will be seen between extremal and nonextremal black holes.

\section{Summary and outlook}
\label{secV}
To summarize, we study the null geodesics in extremal Kerr-Newman black holes.
The roots of the radial potential are classified to find the corresponding parameter regions of the azimuthal angular momentum and the Carter constant of the light rays, which lead to the different types of orbits.
One of the unique features of extremal black holes is the existence of the stable double root at the horizon giving the stable spherical motion.
For the black hole's spin $a<M/2$, in addition to the unstable double root, the stable double root is isolated from the unstable one.
In contrast, for $a\ge M/2$, two types of double roots merge at the triple root and the unstable double root can be at the horizon in some parameter region, resulting in a very different shape of the light ring from the one for $a<M/2$.
Later, the analytical expressions of various light orbits, which can reach spatial infinity to be observed, are obtained. 
In particular, some of them, which start from the NHEKN with the parameters giving the unstable double root and the triple root for $ a\ge M/2$, show a simple expression in terms of elementary functions.
For the unstable double root at the horizon, when the initial $r_i$ of the light rays is near the horizon, the coordinate time and azimuthal angle reveal the power-law dependence of $\vert r_i-r_h \vert^{-1/2}$ to reach spatial infinity. For the triple root case, they are like $\vert r_i-r_h \vert^{-3/2}$ instead.
Those infrared divergences when $r_i \rightarrow r_h$ should play a key role in the study of the extreme black hole shadow or orbiting hot spots \cite{gralla-2018}.
Moreover, the analytical solutions for the equatorial motion can also shed light on the deflection of the light from the source to the observers by the black holes.
When changing the azimuthal angular momentum, namely the impact parameter, either the double or triple root at the horizon is approached from the turning point in the SDL.
The power-law divergence in the deflection angle, consistent with the infrared behavior of the orbits as the initial $r_i$ is near the horizon, is found, showing a stronger divergence than that for the corresponding roots in nonextremal black holes.
This might be another interesting effect of the light deflection by extremal black holes.

\appendix
\section{The integrals $G_{\theta}$, $G_{\phi}$, and $G_{t}$}
\label{appA}
We give here a short review of the angular potential and the solutions of the orbits along the $\theta$ direction.
We begin by rewriting the $\Theta$ potential in terms of $u=\cos^2 \theta$ as
\begin{align}
(1-u)\Theta(u)=-a^2u^2+(a^2-\eta-\lambda^2)u+\eta \, .
\end{align}
The roots of $\Theta(\theta)$ are
\be
u_{\pm}=\frac{\Delta_{\theta}\pm\sqrt{\Delta_{\theta}^2+4\,{a}^2\, \eta}}{2{a}^2}\,,\quad\Delta_{\theta}={a}^2-{\eta-\lambda^2}\, . \label{u+-}
\ee
Apparently, for $\eta >0$ and nonzero $\lambda$, ${1>u_+}>0$ is the only positive root that, in turn, gives two roots at $\theta_+=\cos^{-1}\left(-\sqrt{u_+}\right), \theta_-=\cos^{-1}\left(\sqrt{u_+}\right)$.

For a given trajectory, we find the Mino time during which the trajectory along the $\theta$ direction travels from $\theta_i$ to $\theta$, given by
\begin{align}
\tau=G_{\theta}=p (\mathcal{G}_{\theta_+}- \mathcal{G}_{\theta_-}) + \nu_{\theta_i} \left[(-1)^p\mathcal{G}_{\theta}-\mathcal{G}_{\theta_i}\right], \label{G_theta_tau}
\end{align}
where the trajectory passes through the turning point $p$ times and $\nu_{\theta_i}=\pm_{\theta}$.
The function $\mathcal{G}_{\theta}$ can be obtained through the incomplete elliptic integral of the first kind $F(\varphi|k)$ as
\be \label{g_theta}
\mathcal{G}_{\theta}=-\frac{1}{\sqrt{-u_{-}a^2}}F\left(\sin^{-1}\left(\frac{\cos\theta}{\sqrt{u_{+}}}\right) \left|\frac{u_+}{u_-}\right)\right.\;.
\ee
The inversion of (\ref{G_theta_tau}) for $p=1$ gives $\theta(\tau)$ as
\be \label{theta_tau}
\theta(\tau)=\cos^{-1}\left(-\nu_{\theta_i}\sqrt{u_+}{\rm sn}\left(\sqrt{-u_{-}a^2}\left(\tau+\nu_{\theta_i}\mathcal{G}_{\theta_i}\right)\left|\frac{u_+}{u_-}\right)\right.\right)
\ee
involving the Jacobi elliptic sine function ${\rm sn}(\varphi|k)$.
Here we have set $\tau_i=0$. The other relevant integrals are given by
\begin{align}
&G_{\phi}(\tau)=\frac{1}{\sqrt{-u_{-}a^2}}\Pi\left(u_{+};{\rm am}\left(\sqrt{-u_{-}a^2}\left(\tau+\nu_{\theta_i}\mathcal{G}_{\theta_i}\right)\left|\frac{u_+}{u_-}\right)\right.\left|\frac{u_+}{u_-}\right)\right.-\nu_{\theta_i}\mathcal{G}_{\phi_i}\label{G_phi_tau}\;,\\
&\mathcal{G}_{\phi_i}=-\frac{1}{\sqrt{-u_{-}a^2}}\Pi\left(u_{+};\sin^{-1}\left(\frac{\cos\theta_i}{\sqrt{u_{+}}}\right)\left|\frac{u_+}{u_-}\right)\right.\label{g_phi}\;,\\
&G_{t}(\tau)=-\frac{2u_{+}}{\sqrt{-u_{-}a^2}}E'\left({\rm am}\left(\sqrt{-u_{-}a^2}\left(\tau+\nu_{\theta_i}\mathcal{G}_{\theta_i}\right)\left|\frac{u_+}{u_-}\right)\right.\left|\frac{u_+}{u_-}\right)\right.-\nu_{\theta_i}\mathcal{G}_{t_i}\label{G_t_tau}\;,\\
&\mathcal{G}_{t_i}=\frac{2u_{+}}{\sqrt{-u_{-}a^2}}E'\left(\sin^{-1}\left(\frac
{\cos\theta_i}{\sqrt{u_{+}}}\right)\left|\frac{u_+}{u_-}\right)\right. \;, \label{g_t}
\end{align}
where the incomplete elliptic integral of the second $E(\varphi|k)$ and third kinds $\Pi(n;\varphi|k)$ are also involved.
We also need the formula of the derivative
\begin{align}
E'\left(\varphi\left|k\right)\right.=\partial_k E\left(\varphi\left|k\right)\right.=\frac{E\left(\varphi\left|k\right)\right.-F\left(\varphi\left|k\right)\right.}{2 k} \,.
\end{align}

\section{The roots of the radial potential and the integrals $I_{\phi}$ and $I_{t}$}
\label{appB}
In this appendix, we first summarize the solutions for the roots of the radial potential, where $R(r)$ can be rewritten as a quartic function
\begin{align}
R(r)=r^4+Ur^2+Vr+W
\end{align}
with the coefficient functions given by
\begin{align}
&U=a^2-\eta-\lambda^2\;,\\
&V=2M\left[\eta+\left(\lambda-a\right)^2\right]\;,\\
&W=- a^2\eta - Q^2\left[\eta+\left(\lambda-a\right)^2\right]\;.
\end{align}
There are four roots, namely, $R(r)=(r-r_1)(r-r_2) (r-r_3) (r-r_4)$ with the property $r_{1}+r_{2}+r_{3}+r_{4}=0$, and they can be written as
\begin{align}
r_{1}&=-z-\sqrt{-\hspace*{1mm}\frac{U}{2}-z^2+\frac{V}{4z}}\;,\\
r_{2}&=-z+\sqrt{-\hspace*{1mm}\frac{U}{2}-z^2+\frac{V}{4z}}\;,\\
r_{3}&=+z-\sqrt{-\hspace*{1mm}\frac{U}{2}-z^2-\frac{V}{4z}}\;,\\
r_{4}&=+z+\sqrt{-\hspace*{1mm}\frac{U}{2}-z^2-\frac{V}{4z}}\;,
\end{align}
where the following notation has been used:
\begin{align}
z&=\sqrt{\frac{\Omega_{+}+\Omega_{-}-\frac{U}{3}}{2}}\, , \quad \quad \Omega_{\pm}=\sqrt[3]{-\hspace*{1mm}\frac{\varkappa}{2}\pm\sqrt{\left(\frac{\varpi}{3}\right)^3+\left(\frac{\varkappa}{2}\right)^2}} \;,
\end{align}
with
\begin{align}
\mathcal{\varpi}=-\hspace*{1mm}\frac{U^2}{12}-W \, , \quad\quad
\mathcal{\varkappa}=-\hspace*{1mm}\frac{U}{3}\left[\left(\frac{U}{6}\right)^2-W\right]-\hspace*{1mm}\frac{V^2}{8}\,.
\end{align}
Moreover the solutions that involve the integral of the radial potential $R$ are defined as
\be
I_n\equiv \int_{r_i}^{r}r^n\sqrt{\frac{1}{R(r)}}dr, \quad n=1,2 \, , \label{Ing}
\ee
\be
I_\pm \equiv \int_{r_i}^{r}\frac{1}{r-r_\pm}\sqrt{\frac{1}{R(r)}}dr \,, \quad
I_h \equiv \int_{r_i}^{r}\frac{1}{r-r_h}\sqrt{\frac{1}{R(r)}}dr\, ,\quad
\delta I_h = \int_{r_i}^{r}\frac{1}{\left(r-r_h\right)^2}\frac{1}{\sqrt{R(r)}}dr. \label{I+-g}
\ee
In terms of (\ref{Ing}) and (\ref{I+-g}), we can rewrite (\ref{phi}) and (\ref{t}) as
\be
I_\phi(\tau)=\frac{2Ma}{r_+-r_-}\left[\left(r_+-\frac{a \lambda+M^2-a^2}{2M}\right)I_+(\tau)-\left(r_--\frac{a \lambda+M^2-a^2}{2M}\right)I_-(\tau)\right]
\label{I phi by 1 2}
\ee
\ba
I_t(\tau)=& \frac{2Ma}{r_+-r_-}\left[\left(r_+-\frac{M^2-a^2}{2M}\right)\left(r_+-\frac{a \lambda+M^2-a^2}{2M}\right)I_+(\tau)-\left(r_- -\frac{M^2-a^2}{2M}\right)\left(r_- -\frac{a \lambda+M^2-a^2}{2M}\right)I_-(\tau)\right] \nonumber\\
& +\left(2M\right)I_1(\tau)+I_2(\tau)+\left[3M^2+a^2\right]\tau .
\label{I t by 1 2}
\ea

\section{Analytical solutions for the light orbits with the parameters giving two complex-conjugated roots and two real-valued roots of the radial potential}
\label{appC}
Here we consider the parameters mainly in the regime with the point F1 or F2 shown in Fig. \ref{eta_lambda_2} for both $a\ge M/2$ and $a<M/2$, in which the roots of the radial potential have a complex-conjugated pair $r_{3}^*={r}_{4}$ and two real-valued roots $r_i>r_h>r_{2}>r_{1}$.
This means that there is no turning point in the black hole exterior.
The light rays starting from $r_i >r_h$ will plunge directly into the black hole or escape to spatial infinity that have been taken into account in \cite{ogasawara-2020} to compute the escape probabilities.
The analytical solutions of the orbits have been shown in \cite{gralla-2020} for light rays and in \cite{ko-2024} for a particle traveling around nonextremal black holes.
Here we adapt those solutions to the extremal cases.
We write the pair of the complex-conjugate roots in terms of their real and imaginary parts $r_3=b-ia$ and $r_4=b+ia$.
The analytical solution can be expressed as \cite{gralla-2020, ko-2024}
\begin{align}
    \tau^{Ub}(r)=\frac{1}{\sqrt{AB}}F\left(\cos^{-1}{\left(\frac{A \left(r-r_1\right)-B\left(r-r_2\right)}{A \left(r-r_1\right)+B\left(r-r_2\right)}\right)}|k\right)-\mathcal{I}^{Ub}_{0i}
\end{align}
\begin{align}
    r^{Ub}(\tau)=\frac{\left(Br_2-Ar_1\right)+\left(Br_2+Ar_1\right)cn\left(X(\tau)\big|k\right)}{\left(B-A\right)+\left(B+A\right)cn\left(X(\tau)\big|k\right)}\, ,
\end{align}
where
\begin{align}
    &A^2=a^2+\left(b-r_2\right)^2>0\, , \notag\\
    &B^2=a^2+\left(b-r_1\right)^2>0\, ,\notag\\
    &k=\frac{\left(A+B\right)^2-\left(r_2-r_1\right)^2}{4AB}\, , \notag\\
    &X^{Ub}(\tau)=\sqrt{AB}\left(\tau+\nu_r\mathcal{I}^{Ub}_{0i}\right)\, .\notag
\end{align}
Notice that $\rm{cn}$ is the Jacobi cosine function and $A>B>0$, $0<k<1$.
${I^{Ub}_\phi}$ and $I^{Ub}_t$ reduce to the forms as in (\ref{I_phi_4}) and (\ref{I_t_4}) where
$I^{Ub}_{1}(\tau)$, $I^{Ub}_{2}(\tau)$ and $I_{h}(\tau)$ are obtained as
\begin{align}
&I^{Ub}_{1}(\tau)=\frac{Br_2+Ar_1}{B+A}\frac{X^{Ub}(\tau)}{\sqrt{AB}}+\frac{2(r_2-r_1)\sqrt{AB}}{B^{2}-A^{2}}R_{1}(\alpha;\Upsilon^{Ub}_{\tau}|k)-\mathcal{I}^{Ub}_{1i}\\
&I^{Ub}_{2}(\tau)=\left(\frac{Br_2+Ar_1}{B+A}\right)^{2}\frac{X^{Ub}(\tau)}{\sqrt{AB}}\left.+4\left(\frac{Br_2+Ar_1}{B+A}\right)\frac{(r_2-r_1)\sqrt{AB}}{B^{2}-A^{2}}R_{1}(\alpha;\Upsilon^{Ub}_{\tau}|k)\right.\notag\\
&\quad\quad\quad\quad+\sqrt{AB}\left(\frac{2(r_2-r_1)\sqrt{AB}}{B^{2}-A^{2}}\right)^{2}R_{2}(\alpha;\Upsilon^{Ub}_{\tau}|k)-\mathcal{I}_{2i}^{Ub}\\
&I^{Ub}_{h}(\tau)=-\frac{1}{B\left(r_{h}-r_{2}\right)+A \left(r_{h}-r_{1}\right)}\left[\frac{B+A}{\sqrt{AB}}X^{Ub}(\tau)
+\frac{2(r_{2}-r_{1})\sqrt{AB}}{B\left(r_{h}-r_{2}\right)-A \left(r_{h}-r_{1}\right)}R_{1}(\alpha_{h};\Upsilon^{Ub}_{\tau}|k) \right]-\mathcal{I}^{Ub}_{hi}
\end{align}
together with $\delta I^{Ub}_h$
\begin{align}
&\delta I^{Ub}_h(\tau)=
\frac{1}{\left[B (r_h - r_1) + A (r_h - r_2)\right]^2} \left[\frac{(A + B)^2 X^{Ub}(\tau)}{\sqrt{AB}} + \frac{2 \sqrt{AB} (r_2 - r_1) (A + B)}{B (r_h - r_1) - A (r_h - r_2)} R_{1}(\alpha_{h};\Upsilon^{Ub}_{\tau}|k)\right]
\notag\\
&\quad\quad\quad
+ \frac{4 (AB)^{3/2} (r_2 - r_1)^2}{\left[B^2 (r_h - r_1)^2 - A^2 (r_h - r_2)^2\right]^2} R_{2}(\alpha_{h};\Upsilon^{Ub}_{\tau}|k)-\mathcal{I}^{Ub}_{dhi}\, .
\end{align}
The parameters of the elliptical integrals given above are defined to be
\begin{align}
    \alpha_h=\frac{B(r_{h}-r_{2})+A(r_{h}-r_{1})}{B(r_{h}-r_{2})-A(r_{h}-r_{1})},\quad\quad\quad
    \alpha=\frac{B+A}{B-A}, \quad\quad\quad
    \Upsilon^{Ub}_{\tau}={\rm am}\left(X(\tau)\left|k\right)\right.\, .
\end{align}
Moreover, $\rm{am}$ is the Jacobi amplitude function and $R_1$ and $R_2$ are the integrals of the $\rm{cn}$ function
\be
R_{1}(\alpha;\phi|k)\equiv\int_{0}^{F(\phi|k)}\frac{du}{1+\alpha {\rm cn}(u|k)}=\frac{1}{1-\alpha^2}\left[\Pi\Bigg(\frac{\alpha^2}{\alpha^2-1};\phi\left|k\Bigg)\right.-\alpha f(p_\alpha,\phi,k)\right] \label{R1}
\ee
\begin{align}
&R_{2}(\alpha;\phi|k)\equiv\int_{0}^{F(\phi|k)}\frac{du}{[1+\alpha {\rm cn}(u|k)]^2}\notag\\
&\quad\quad=\frac{1}{\alpha^2-1}\left[F\left(\phi|k\right)-\frac{\alpha^2}{k+(1-k)\alpha^2}\left(E(\phi|k)-\frac{\alpha\sin(\phi)\sqrt{1-k\sin^2(\phi)}}{1+\alpha\cos(\phi)}\right)\right]\notag\\
&\quad\quad\quad+\frac{1}{k+(1-k)\alpha^2}\left(2k-\frac{\alpha^2}{\alpha^2-1}\right)R_{1}(\alpha;\phi|k) \label{R2}
\end{align}
in which
\begin{align}
&f(p_\alpha,\phi,k)=\frac{p_\alpha}{2}\ln\left(\frac{p_\alpha\sqrt{1-k\sin^2(\phi)}+\sin(\phi)}{p_\alpha\sqrt{1-k\sin^2(\phi)}-\sin(\phi)}\right)\, , \quad p_\alpha=\sqrt{\frac{\alpha^2-1}{k+(1-k)\alpha^2}}\, ,
\end{align}
where for $-1<\alpha<1$, the solution is a real and continuous function.


\begin{acknowledgments}
This work was supported in part by the National Science and Technology Council (NSTC) of Taiwan, Republic of China.
\end{acknowledgments}


\addcontentsline{toc}{chapter}{Bibliography}
\bibliography{References}

\end{document}